\documentclass[fleqn,usenatbib]{mnras}
\usepackage{import}

\import{./frontmatter/}{preamble.sty}
\def \aj {{\rm {AJ}}}
\def \araa {{\rm {ARA\&A}}}
\def \apj {{\rm {ApJ}}}

\def \apjs {{\rm {ApJS}}}
\def \apjl {{\rm {ApJL}}}

\def \aap {{\rm {A\&A}}}
\def \aapr {{\rm {A\&AR}}}

\def \mnras {{\rm {MNRAS}}}

\def \pasj {{\rm {PASJ}}}
\def \pasp {{\rm {PASP}}}

\def \nat {{\rm {Nat}}}
\def \kms {{\rm km\,s$^{-1}$}}

\def \sol {{\rm M$_\odot$}}
\def \Lsol {{\rm L$_\odot$}}
\def \arcsec {{\rm $^{\prime\prime}$}}
\def \micron{\hbox{$\upmu$m}}

\def \HII {{\rm \ion{H}{II}}}

\newcommand{\note}[1]{\textcolor{black}{  #1}}

\title[Feedback in high-pressure environments]{Which feedback mechanisms dominate in the high-pressure environment of the Central Molecular Zone?}

\author[A.~Barnes, S.~N.~Longmore, et al.]
{Ashley T.~Barnes,$^{1}$\thanks{E-mail: ashleybarnes.astro@gmail.com}
Steven N.~Longmore,$^{2}$  
James E.~Dale,$^{3}$
Mark R.~Krumholz,$^{4,5}$ \and
J.~M.~Diederik Kruijssen$^{6}$ and
Frank Bigiel$^{1}$
 \\ 
$^{1}$Argelander-Institut f\"{u}r Astronomie, Universit\"{a}t Bonn, Auf dem H\"{u}gel 71, 53121, Bonn, DE \\
$^{2}$Astrophysics Research Institute, Liverpool John Moores University, 146 Brownlow Hill, Liverpool L3 5RF, UK \\
$^{3}$Centre for Astrophysics Research, University of Hertfordshire, Hatfield, AL10 9AB, UK\\
$^{4}$Research School of Astronomy and Astrophysics, Australian National University, Canberra, ACT 2611 \\
$^{5}$ARC Centre of Excellence for Astronomy in Three Dimensions (ASTRO-3D), Canberra, ACT 2601 Australia \\
$^{6}$Astronomisches Rechen-Institut, Zentrum f\"{u}r Astronomie der Universit\"{a}t Heidelberg, M\"{o}nchhofstra\ss e 12-14, 69120 Heidelberg, Germany \\
}

\date{Accepted 2020 September 3. Received 2020 September 3; in original form 2020 July 14.}

\pubyear{2020}

\begin{document}

\label{firstpage}
\pagerange{\pageref{firstpage}--\pageref{lastpage}}
\maketitle

\begin{abstract}
 Supernovae (SNe) dominate the energy and momentum budget of stellar feedback, \note{but} the efficiency with which they couple to the interstellar medium (ISM) depends strongly on how effectively early, pre-SN feedback clears dense gas from star-forming regions. There are observational constraints on the magnitudes and timescales of early stellar feedback in low ISM pressure environments, yet no such constraints exist for more cosmologically typical high ISM pressure environments. In this paper, we determine the mechanisms dominating the expansion of \HII\ regions as a function of size-scale and evolutionary time within the high-pressure ($P/k_\mathrm{B}\,\sim\,10^{7-8}$ K\,cm$^{-3}$) environment in the inner 100\,pc of the Milky Way. We calculate the thermal pressure from the warm ionised ($P_\mathrm{HII}$; 10$^{4}$\,K) gas, direct radiation pressure ($P_\mathrm{dir}$), and dust processed radiation pressure ($P_\mathrm{IR}$). We find that (1) $P_\mathrm{dir}$ dominates the expansion on small scales and at early times (0.01-0.1\,pc; $<$0.1\,Myr); (2) the expansion is driven by $P_\mathrm{HII}$ on large scales at later evolutionary stages ($>0.1$\,pc; $>1$\,Myr); (3) during the first $\lesssim 1$ Myr of growth, but not thereafter, either $P_{\rm IR}$ or stellar wind pressure likely make a comparable contribution. Despite the high confining pressure of the environment, natal star-forming gas is efficiently cleared to radii of several pc within $\sim$\,2\,Myr, i.e. before the first SNe explode. This `pre-processing' means that subsequent SNe will explode into low density gas, so their energy and momentum will efficiently couple to the ISM. We find the \HII\ regions expand to a radius of $\sim$\,3pc, at which point they have internal pressures equal with the surrounding external pressure. A comparison with \HII\ regions in lower pressure environments \note{shows that} the maximum size of all \HII\ regions \note{is} set by pressure equilibrium with the ambient ISM.
 \end{abstract}


\section{Introduction}\label{sec:Introduction}

\begin{figure*}
\centering
\includegraphics[trim = 0mm 0mm 0mm 0mm, clip,angle=0,width=1 \textwidth]{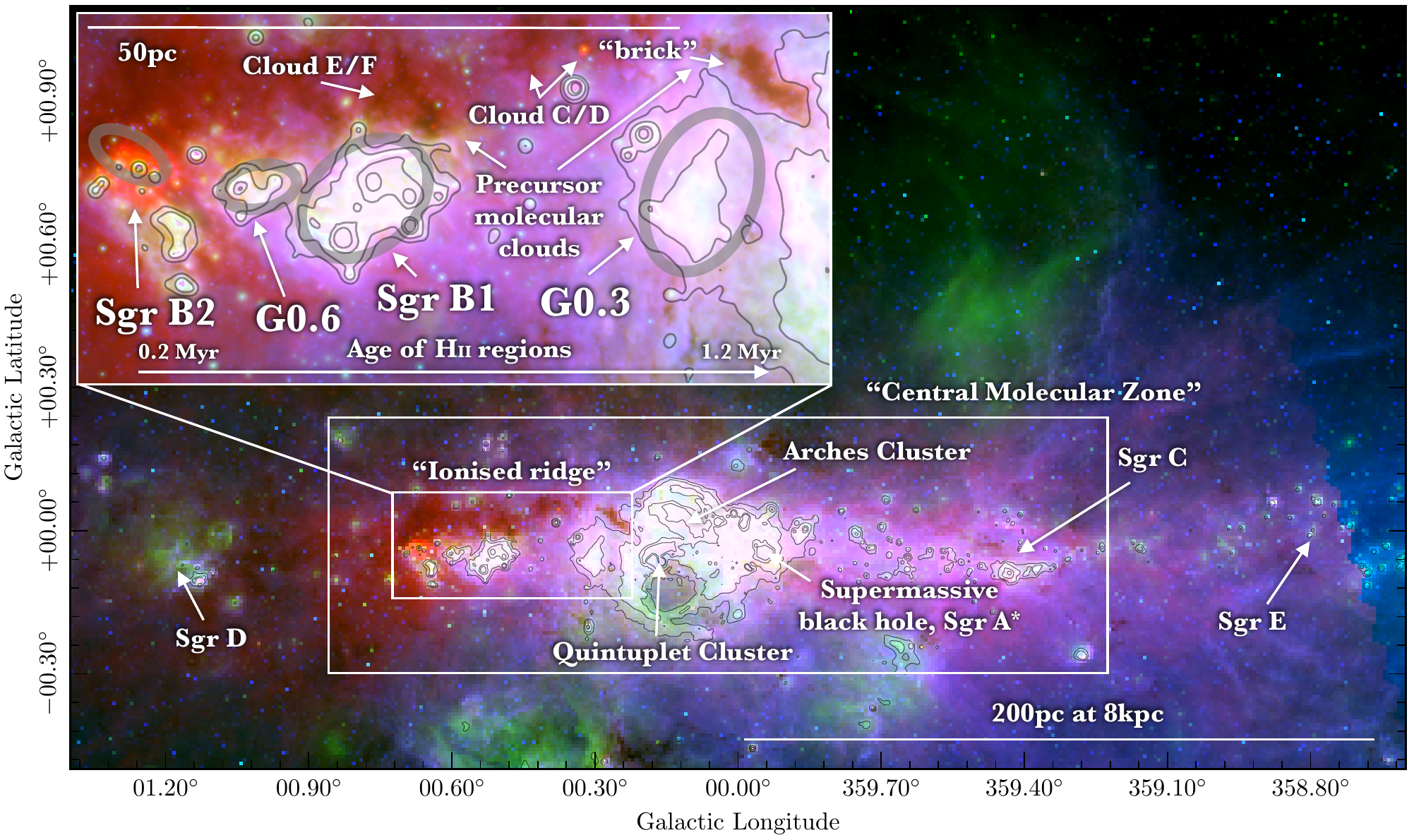}
\caption{A three colour image of the Galactic Centre. In this image, red is 70\,\micron\ emission from {\it Herschel} Hi-GAL \citep{molinari_2010}, green is 24\,\micron\ emission from {\it Spitzer} MIPSGAL \citep{carey_2009}, and blue is 8\,\micron\ emission from {\it Spitzer} GLIMPSE \citep{churchwell_2009}. We label sources of interest throughout this region, such as the prominent massive clusters (Arches and Quintuplet; e.g. \citealp{espinoza_2009, harfst_2010}), the supermassive black hole (Sgr A*; e.g. \citealp{grav_collab_2018}), and several prominent \HII\ regions (e.g. Sgr E; \citealp{anderson_2020}). Rectangles show the approximate regions of the Central Molecular Zone (or CMZ) and the dust-ridge. The 24\,\micron\ emission map is also overlaid in contours of 400, 1000, and 1500 MJy\,sr$^{-1}$, which have been chosen to best highlight the sources of interest. The inset in the upper left shows a zoom-in of the ionised-ridge region, which contains the sources Sgr B2, G0.6, Sgr B1, and G0.3 that we study in this work. In the zoom-in panel, we also label molecular clouds that are thought to represent the initial conditions of the \HII\ regions (see section\,\ref{sec:energetics}). Labels at the bottom of the zoom-in panel give the approximate ages of the \HII\ regions based on the predictions from an orbital model (increasing age from left to right; \citealp{longmore_2013a, kruijssen_2015}). In the lower right of the main panel and upper left of the zoom-in panel we show scale-bars representing projected lengths of $\sim$\,200\,pc and $\sim$\,50\,pc at a distance of $\sim\,8\,$\,kpc, respectively \citep{reid_2014, grav_collab_2018}.}
\label{three_col_map}
\end{figure*}

 Throughout their short lifetimes, high-mass stars ($>8$\,\sol) inject large amounts of energy and momentum into their host environments through a variety of feedback processes (e.g. \citealp{krumholz_2014}). The most potentially disruptive of these feedback mechanisms occurs when the stars eventually die, exploding as supernovae (SNe). Indeed, SNe are thought to play a major role in the self regulation of star formation in galaxies through their contribution to the total energy and momentum budget of the interstellar medium (ISM; \citealp{mckee_1977, maclow_2004, klessen_2016}).

As the rate of cooling in the ISM is proportional to the gas density squared, the efficiency with which SNe inject energy and momentum into the local galactic environment strongly depends on the density distribution of the gas into which they explode (see \citealp{girichidis_2016} and references therein). For example, SNe that explode within dense molecular clouds may be limited to disrupting their natal gas clouds, whilst SNe that explode into lower density environments can drive hot expanding bubbles to much larger distances (tens to hundreds of pc) and influence galactic scale processes (e.g. kpc-scale galactic outflows; \citealp{veilleux_2005, agertz_2013, stinson_2013, keller_2020, veilleux_2020}).

Feedback from the pre-SNe stages of high-mass stars plays a significant role in determining the environment into which SNe subsequently explode. \note{Simulations have long predicted that this} `pre-processing' can potentially even destroy the host molecular cloud before the first SN explosion (e.g. \citealp{dale_2012, dale_2013}), and observations of molecular clouds and \HII\ regions in nearby galaxies now show that pre-SN feedback is primarily responsible for the destruction of molecular clouds across the local galaxy population \citep{kruijssen19a,chevance20b,chevance20}. Studying the effects of these earliest stages of stellar feedback on their environment is then crucial to quantifying the contribution of SNe in driving the \note{galaxy-scale} energy and momentum cycle of the ISM in galaxies.

In light of this, significant observational effort has been invested to better disentangle and quantify the effect of various feedback mechanisms within young stellar systems (e.g. \citealp{oey_1996a, oey_1996b, pellegrini_2010, pellegrini_2011}). More recent efforts have focused on measuring and comparing the internal pressure components from different feedback mechanisms in \HII\ regions located within the Small and Large Magellanic Clouds (SMC and LMC, respectively), such as the well known 30 Doradus complex (e.g. \citealp{lopez_2011, lopez_2014, chevance16, mcleod_2019})\note{, as well as other nearby galaxies \citep[e.g.][]{mcleod_20}}. These studies have provided important insights into early-stage feedback, but further work is needed to understand how  pre-processing varies with environment, particularly to higher density, pressure, and metallicity regimes such as those in galactic nuclei and high-redshift galaxies.

\note{Most of the literature to date naturally focuses on regions that are observationally accessible in the local Universe, which are characterised by low gas pressures ($P/k\sim10^{4{-}5}~{\rm K}~{\rm cm}^{-3}$). The ambient gas pressure plays a major role in the molecular cloud lifecycle \citep{chevance20b}, specifically by setting the initial conditions for star formation \citep[e.g.][]{faesi18,sun18,sun20,jeffreson20}, the subsequent star formation efficiency \citep[e.g.][]{krumholz05,blitz06,federrath12}, and the impact of stellar feedback \citep[e.g.][]{grudic18,kim18,fujimoto19,li19,keller_2020}. This is particularly important, because ISM pressures observed at the peak of the cosmic star formation history are several orders of magnitude higher than those observed in disc galaxies today \citep[e.g.][]{genzel11,swinbank11,swinbank12,tacconi13}. It is therefore a critical question how the physics of star formation and feedback proceeded under the extreme pressures observed at the time that the Universe was forming stars most rapidly.}

In this work, we investigate the effect of higher ambient density and pressure host environments on the physical properties and evolution of pre-SNe high-mass star formation regions. For this, we focus on the inner few hundred parsecs of the Milky Way, known as the ``Central Molecular Zone'' (CMZ; e.g. \citealp{morris_1996}), which is known to host several (pre-SNe) \HII\ region complexes (e.g. \citealp{mehringer_1992, schmiedeke_2016, anderson_2020}). This region has average gas densities, gas temperatures, ambient pressures, turbulent velocity dispersions, interstellar radiation fields and cosmic ray ionisation rates factors of a few to several orders of magnitude larger than observed in typical Milky Way disc and SMC/LMC star-forming systems, more similar to starburst and high redshift galaxies at the epoch of peak star formation density at z\,$\sim\,1-3$ \citep{kruijssen_2013}. 

Figure\,\ref{three_col_map} presents a three-colour image of the Galactic Centre with the CMZ region highlighted by the white rectangle, and several sources of interest within the regions labelled, such as the Arches and Quintuplet young massive clusters (e.g. \citealp{espinoza_2009, harfst_2010}) and the central supermassive black hole (Sgr A$^*$; e.g. \citealp{grav_collab_2018}). The inset image shows a zoom-in of the so-called ``ionised ridge'' region of the CMZ that contains the \HII\ regions complexes we study this work: Sgr B2, G0.6, Sgr B1, and G0.3. We also highlight the quiescent precursor molecular clouds on the ``dust-ridge'' (e.g. \citealp{lis_1994, longmore_2012, longmore_2013, walker_2015, walker_2016, walker_2018, barnes_2019,  henshaw_2019, battersby_2020}). This dust-ridge is thought to be connected to the ionised ridge via a ring of material surrounding the Galactic Centre found in simulations (e.g. \citealp{sormani_2018,dale_2019,kruijssen19b, sormani20,tress20}) and extragalactic systems (e.g. \citealp{comeron_2010, krieger_2020}), which is maintained by the inflow of material from the bar (e.g. \citealp{krumholz_2015, henshaw_2016e,  sormani_2019}). On the Figure\,\ref{three_col_map} inset, we also show age estimates of the \HII\ regions based on the predictions from an orbital model (increasing age from left to right; \citealp{longmore_2013a, kruijssen_2015}). We summarise the physical properties of the precursor molecular clouds and the \HII\ regions in Table\,\ref{source_properties} (adopted from \citealp{barnes_2017}). 

This paper is organised as follows. In section\,\ref{sec_results} we outline how we determine each of the internal pressure components of the Galactic Centre \HII\ regions. In section\,\ref{sec_analysis} we analyse the radial dependence of these pressure components and estimate the energies and momenta of the expanding \HII\ regions. In section\,\ref{sec_discussion} we discuss our findings in the context of the previous observations of the SMC and LMC, compare to analytic models for the expansion of \HII\ regions, and explore how efficiently the stellar population is driving the expansion. Finally, we summarise the results in section\,\ref{sec_conclusions}.

\begin{table}
\centering
\caption{The global properties of the Galactic Centre molecular clouds and \HII\ regions \citep[][Tables 4 and 5]{barnes_2017}. The columns list the effective radii, molecular gas masses, total embedded stellar mass, the times since pericentre passage (i.e. point of triggered molecular clouds collapse) as estimated from the orbital model of \citet{kruijssen_2015}, and the age of the \HII\ regions defined as $t _\mathrm{p,last} - t _\mathrm{p,last} \mathrm{(dust-ridge)}$ (section\,\ref{sec:energetics}). In this work, we make the assumption that the dust-ridge molecular clouds represent the precursors to the identified \HII\ regions (see Figure\,\ref{three_col_map}).}
\begin{tabular}{c c c c c c}
\hline
Source & $R_\mathrm{eff}$ & $M_\mathrm{gas}$ & $M_\mathrm{*}$ & $t _\mathrm{p,last}$ & $t _\mathrm{age}$ \\
& (pc)  & ($10^4$\,\sol) & ($10^3$\,\sol) & (Myr) & (Myr)  \\
\hline
Dust-ridge$^{*}$ & 1.6 & 3.7 & <0.1 & <0.53 & 0 \\
Sgr B2 & 2.7 & 65 & 33 & 0.74 & 0.21  \\
G0.6 & 2.8 & 4.6 & 3.3 & 1.45 & 0.92 \\
Sgr B1 & 5.8 & 8.7 & 8.0 & 1.55 & 1.02 \\ 
G0.3 & 6.5 & 9.3 & 6.2 & 1.75 & 1.22 \\ 
\hline
\end{tabular}
{\vspace{0.3cm}}
\begin{minipage}{0.9\columnwidth}
\vspace{1mm}
$^*$Shown is the average radius, gas mass, and stellar mass of the ``Brick'', ``b'', ``c'', ``d'', ``e'' and ``f'' molecular clouds on the dust-ridge \citep{barnes_2017}.\\
\end{minipage}
\label{source_properties}
\end{table}
\section{Pressure calculation}\label{sec_results}

\subsection{Internal \HII\ region pressure components}\label{sec:Pressure}

The internal pressure within an \HII\ region produced by an embedded stellar population can be expressed as the sum of four dominant pressure components (e.g. \citealp{lopez_2011, lopez_2014, mcleod_2019}):
\begin{itemize}
\item [$P_\mathrm{HII}$)] warm ionised thermal gas pressure,
\item [$P_\mathrm{dir}$)] direct radiation pressure, 
\item [$P_\mathrm{IR}$)] dust processed radiation pressure, 
\item [$P_\mathrm{X}$)] hot X-ray emitting thermal gas pressure,
\end{itemize}
such that the total internal pressure is, 
\begin{equation}
P_\mathrm{tot} = P_\mathrm{HII} + P_\mathrm{dir} + P_\mathrm{IR} + P_\mathrm{X}. 
\end{equation}

In this section, we discuss the methodology used to determine the $P_\mathrm{HII}$, $P_\mathrm{dir}$, and $P_\mathrm{IR}$ pressure components within the Galactic Centre \HII\ regions. We mention $P_\mathrm{X}$ here because it is included in the comparison to the LMC and SMC (section\,\ref{sec:comp2SMCLMC}; \citealp{lopez_2011, lopez_2014}). Lopez et al. determine $P_\mathrm{X}$ by modelling the thermal (free-free) emission from the hot gas, which peaks at soft X-ray energies for temperatures of $T_\mathrm{x}\sim$\,10$^{6}$\,K ($k_\mathrm{B}T_\mathrm{x}\sim$\,0.1\,keV). Unfortunately, we cannot make a similar measurement for the Galactic Centre region, for two reasons. First, the Galactic Centre hosts significant non-thermal soft X-ray emission that is expected to be significantly brighter than any potential thermal emission (e.g. \citealp{law_2005, yusef_2007, nobukawa_2008, ponti_2015, zhang_2015}). Second, the substantial column density of foreground material towards the Galactic Centre \HII\ regions is expected to make any thermal soft X-ray emission undetectable. For example, the average molecular hydrogen column density for the \HII\ regions is $\sim\,3\,\times\,10^{22}$cm$^{-2}$, which when using the dust grain models from \citet{draine_2003} corresponds to a line-of-sight extinction of $A_\mathrm{100eV}>20$\,mag.\footnote{Using $A_\lambda = (2.5/\ln 10)\mathrm{C_{ext}(\lambda)\,N_H} = 1.086\mathrm{C_{ext}(\lambda)\,N_H}$, and the extinction cross section per H nucleon, $\mathrm{C_{ext}(\lambda)}$, at 100\,eV of $\mathrm{C_{ext}(0.0124\,\mu\,m)}=7.238\,\times\,10^{-22}$\,cm$^2$ \note{(see Table\,6 of \citealp{draine_2003})}.} Such a high X-ray extinction would cause even the most luminous known \HII\ regions to become undetectable for \note{any} plausible amount of e.g. {\it Chandra} observing time. We, therefore, omit the determination of $P_\mathrm{X}$ \note{within the Galactic Centre} from our analysis.

\subsubsection{$P_\mathrm{HII}$: Thermal pressure from warm (10$^4$\,K) ionised gas}\label{subsec:Warm ionised gas}

\begin{table*}
\centering
\caption{The sample of radio observations taken from the literature that have been used to calculate the warm ionised gas pressure. We list the telescope, the frequency, and reference for each observation.}
\begin{tabular}{c c c c}
\hline
ID & Telescope & Frequency (GHz) & References \\
\hline

1 & Pencil-beam antenna (Single dish) & 2  \& 4  \& 6  \& 10 & \citet{downes_1966} \\
2 & NRAO's 36-foot reflector (Single dish) & 31 & \citet{downes_1970} \\
3 & Naval Research Laboratory's 85-foot reflector (Single dish) & 11 \& 18 \& 32 & \citet{hobbs_1971a} \\
4 & NRAO's 36-foot reflector (Single dish) & 85 & \citet{hobbs_1971b} \\
5 & Wilkinson Microwave Anisotropy Probe (WMAP; Single dish) & 93 & \citet{lee_2012} \\
6 & Molonglo Observatory Synthesis Telescope (MOST; Interferometer) & 0.4 & \citet{little_1974} \\
7 & Very Large Array (VLA; Interferometer) & 5 \& 15 & \citet{benson_1984} \\
8 & Very Large Array (VLA;Interferometer) & 15 & \citet{roelfsema_1987} \\
9 & Very Large Array (VLA;Interferometer) & 15 & \citet{gaume_1990} \\
10 & Very Large Array (VLA;Interferometer) & 1.5 \& 5 \& 8 & \citet{mehringer_1992} \\
11 & Very Large Array (VLA;Interferometer) & 0.3 & \citet{yusef_2007} \\
12 & Very Large Array (VLA;Interferometer) & 23 \& 43 & \citet{schmiedeke_2016} \\

\hline
\end{tabular}
\label{radio_obs}
\end{table*}

\HII\ regions are ionised by the large flux of H-ionizing Lyman continuum, $\mathcal{N}_\mathrm{LyC}$, produced by young high-mass stars ($>$8\sol). This pressure in this ionized gas is set by the ideal gas law, $P_\mathrm{HII}\,=\,2\,n_\mathrm{e} k_\mathrm{B}\,T_\mathrm{HII}$, where the factor of two comes from the assumption that all the He is singly ionised (i.e. $n_\mathrm{e} + n_\mathrm{H} + n_\mathrm{He} = 2n_\mathrm{e}$). Free-free interactions between the electrons and ions gives rise to continuum emission at cm-(mm) wavelengths. The intensity and physical size of this emission can be used to derive the electron density of the ionised gas.

\citet{lopez_2011, lopez_2014} determine the electron density for the LMC and SMC sources from their radio continuum fluxes using the expression given by \citet{rybicki_1979}. However, we choose to use the conversion presented by \citet{mezger_1967}, as this was adopted by the most comprehensive survey of (ultra-compact) \ion{H}{II} regions within the Galactic Centre sample by \citet{schmiedeke_2016}. The difference in derived electron density using the two methods is only of order 25 per cent, so does not affect the conclusions drawn from the data.\footnote{This comparison was made for a 1\,Jy source at 5\,GHz, with a radius of 1\arcsec\ at 8\,kpc. For this calculation we use equation 5.14b from \citet{rybicki_1979}, rather than \citet[][equ. 6]{lopez_2011} and \citet[][equ. 10]{lopez_2014}, which have an incorrect constant of 6.8$\times10^{38}$\,cm$^{-3}$\,erg$^{-1}$\,K$^{-0.5}$. Rearranging equation 5.14b from \citet{rybicki_1979}, we find a constant of 1.46$\times10^{37}$\,cm$^{-3}$\,erg$^{-1}$\,K$^{-0.5}$.} The \citet{mezger_1967} conversion can be given as, 
\begin{multline}
n_\mathrm{e} = 2.576\,\times\,10^{6} \left( \frac{F_\nu}{\mathrm{Jy}} \right)^{0.5}  \left(\frac{T_\mathrm{HII}}{\mathrm{K}}\right)^{0.175} \left(\frac{\nu}{\mathrm{GHz}}\right)^{0.05} \\ \left(\frac{\theta_\mathrm{source}}{\mathrm{arcsec}}\right)^{-1.5} \left(\frac{D}{\mathrm{pc}}\right)^{-0.5} \mathrm{cm^{-3}},
\label{mh}
\end{multline}
where $F_\nu$ is the measured flux, $\nu$ is the frequency of the observations, $T_\mathrm{HII}$ is the electron temperature, $D$ is the source distance and $\theta_\mathrm{source}$ is the circular diameter of the source.  

Observations of the Galactic Centre at radio wavelengths have been possible for several decades, hence there is an extensive library of data taken at various frequencies and resolutions. We choose a sample of observations from both single-dish and interferometer telescopes, which are listed in Table\,\ref{radio_obs}, along with the frequencies and reference for each observation. To calculate $n_\mathrm{e}$ for each of the literature catalogue flux and diameters, we assume a source distance of 8\,kpc (e.g. \citealp{reid_2016}), and $T_\mathrm{HII}=5000$\,K. These values of $n_\mathrm{e}$ are then used with $T_\mathrm{HII}$ to determine $P_\mathrm{HII}$. 

\note{The low $T_\mathrm{HII}$ used in the above calculation has been chosen to accounts for the well-studied systematic decrease of $T_\mathrm{HII}$ observed at decreasing galactocentric radius, caused by the corresponding increase in metallicity; the typical electron temperature at solar metallicity is $\sim$\,7000\,K (e.g \citealp{mezger_1979, shaver_1983, wink_1983, caswell_1987, deharveng_2000, giveon_2002}). When extrapolating the electron temperature to galactocentric radius relation from \citet{deharveng_2000} down to $R_\mathrm{GC}$\,=\,0\,kpc, we find $T_\mathrm{HII}$\,=\,4260\,$\pm$\,350\,K. This result is in agreement with the median $T_\mathrm{HII}\sim$\,5000\,K measured across a several Galactic Centre \HII\ regions (\citealp{gaume_1990, mehringer_1992, cram_1996, lang_1997, lang_2001}). Therefore, we adopt $T_\mathrm{HII}=5000$\,K as the representative value for Galactic Centre \HII\ regions, and use this electron temperature for all calculations within this work.}

\subsubsection{$P_\mathrm{dir}$: Direct radiation pressure}\label{subsec:Direct radiation}

The large radiation field directly produced from young stellar objects can exert a significant pressure on the surrounding material. This radiation pressure, $P_\mathrm{rad}$, at a given position within an \HII\ region, is related to the bolometric luminosity, $L_\mathrm{bol}$, of the stellar population and the distance, $r$, from each star to that position within the region:
\begin{equation}
P_\mathrm{rad} = \sum{\frac{L_\mathrm{bol}}{4 \pi r^{2} c}},
\label{eq_Prad}
\end{equation}
where the summation is over all stars within the region. The volume-averaged direct radiation pressure, $P_\mathrm{dir}$, is then given as \citep{lopez_2014}, 
\begin{equation}
P_\mathrm{dir} = \frac{3 L_\mathrm{bol}}{4 \pi R^{2} c},
\label{eq_Pdir}
\end{equation}
where $R$ is the radius of the \HII\ region (or effective radius, $R_\mathrm{eff}$, that we define later and use throughout the rest of the paper), and $L_\mathrm{bol}$ is the bolometric luminosity from the population of massive stars within the \HII\ region. This form differs by a factor of three from \citet[][equation 4]{mcleod_2019}, as these authors calculate the radiation surface pressure rather than the volume average pressure. This expression is appropriate to compute the force balance at the surface of an empty shell. However, as this work aims at understanding the large-scale dynamics of the Galactic Centre \HII\ regions (e.g. the total energy and pressure budget for each source), the inclusion of a factor of three in the numerator of the above equation is required. \note{We also note here that the higher metallicity within the Galactic Centre, or increasing the amount of dust, has no effect on the $P_\mathrm{dir}$ calculation. Direct radiation pressure is limited by the momentum supplied by the stellar radiation field, and, as long as there is enough dust around to absorb all the radiation, the momentum per unit time delivered is the same.}

\citet{lopez_2011} determine the bolometric luminosity of the \HII\ regions within the LMC and SMC from H$\upalpha$ emission. However, this is not possible for the \HII\ regions investigated here, due to the high optical extinction towards the Galactic Centre ($A_\mathrm{v}>$20\,mag), which completely obscures any H$\upalpha$ emission. We, therefore, adopt two alternative methods of calculating the bolometric luminosity using radio and infrared observations (i.e. wavelengths where the emission is much less affected by dust extinction). 

Firstly, we can make the assumption that the bolometric luminosity is proportional to the flux of ionising photons, $\mathcal{N}_\mathrm{LyC}$, such that $L_\mathrm{bol} = \mathcal{N}_\mathrm{LyC} \left \langle h \nu \right \rangle$, where $\left \langle h \nu \right \rangle \sim$\,15\,eV is the mean photon energy \citep{pellegrini_2007}. We use the $\mathcal{N}_\mathrm{LyC}$ for each \HII\ region as determined from the radio observations outlined in Table\,\ref{radio_obs} (i.e. \citealp{gaume_1990, mehringer_1992, schmiedeke_2016}), and solve for the direct radiation pressure using equation\,\ref{eq_Pdir}.

The second method assumes that the luminosity integrated over infrared wavelengths approximately corresponds to the total bolometric luminosity. This is a common assumption made for embedded star-forming regions, where the luminosity from massive stars produced at ultraviolet wavelengths is absorbed and remitted by the dust in the infrared. \citet{barnes_2017} have produced maps of the total infrared luminosity across the Galactic Centre. These authors fit a two-component modified black-body function to extinction corrected 5.8\,-\,24\,\micron\ \citep{churchwell_2009, carey_2009} and 160\,-\,500\,\micron\ \citep{molinari_2010} emission maps (referred to as the warm and cool component of the bolometric luminosity; see Fig.\,2 of \citealp{barnes_2017}). These infrared (i.e. bolometric) luminosity maps are used with the two methods outlined below to also determine the direct radiation pressure within each of the Galactic Centre \HII\ regions.

In comparison with the first method for calculating the direct radiation pressure from radio observations, we choose to identify individual sources within the infrared maps. These can be considered as discrete \HII\ regions each with a single value of the direct radiation pressure. We choose to identify these \HII\ regions in the map of the warm component of the bolometric luminosity using a dendrogram analysis \citep{rosolowsky_2008}. We choose to use a structure finding algorithm, as opposed to by-eye identification, to give reproducibility within regions with particularly complex morphology (the warm bolometric luminosity map is given in Fig. 2 from \citealp{barnes_2017}).\footnote{The following set of parameters are used for determination of the dendrogram structure: {\sc min\_value} = 3\,$\sigma$ $\sim$ 300\,\Lsol\ (the minimum luminosity considered in the analysis); {\sc min\_delta} = 3\,$\sigma$ (the minimum spacing between isocontours); {\sc min\_delta} = 1/3\,beam $\sim$ 3 pixels (the minimum number of pixels contained within a structure). Varying the dendrogram parameters over a wide range of values does not affect the results of the paper.} We make use of the ``leaves'' identified from the dendrogram analysis, which are the highest level (i.e. smallest) structures in the analysis and here represent distinct \HII\ regions. We take the mask of each \HII\ regions (dendrogram leaf), and apply this to both the warm and cool bolometric luminosity component maps, which we sum to then get the total bolometric luminosity. This is used with equation\,\ref{eq_Pdir} to get the direct radiation pressure ($P_\mathrm{dir}$) within each \HII\ region. The effective radius ($R_\mathrm{eff}$) of each \HII\ region is defined as the radius for a circle with the corresponding area ($A$) of each structure (i.e. $R_{\rm eff}=\sqrt{A/\pi}$). 

In addition to the dendrogram analysis, we also calculate the direct radiation pressure within apertures of increasing radius from the centre of each \HII\ region complex. To do so, we place circular masks for each source on to both the warm and cool bolometric luminosity component maps, and sum the enclosed values to then get the total bolometric luminosity. The circle is then increased in radius, and the process repeated. We again use equation\,\ref{eq_Pdir} to determine the direct radiation pressure within these increasing circular apertures. This method differs from the dendrogram analysis, as it returns a continuous radial distribution from the source centre, as opposed to a distribution of distinct \HII\ region with various sizes.

\subsubsection{$P_\mathrm{IR}$: Dust-processed radiation pressure}\label{subsec:Dust reprocessed emission}

\begin{figure}
\centering
\includegraphics[trim = 0mm 0mm 0mm 0mm, clip,angle=0, width=1\columnwidth]{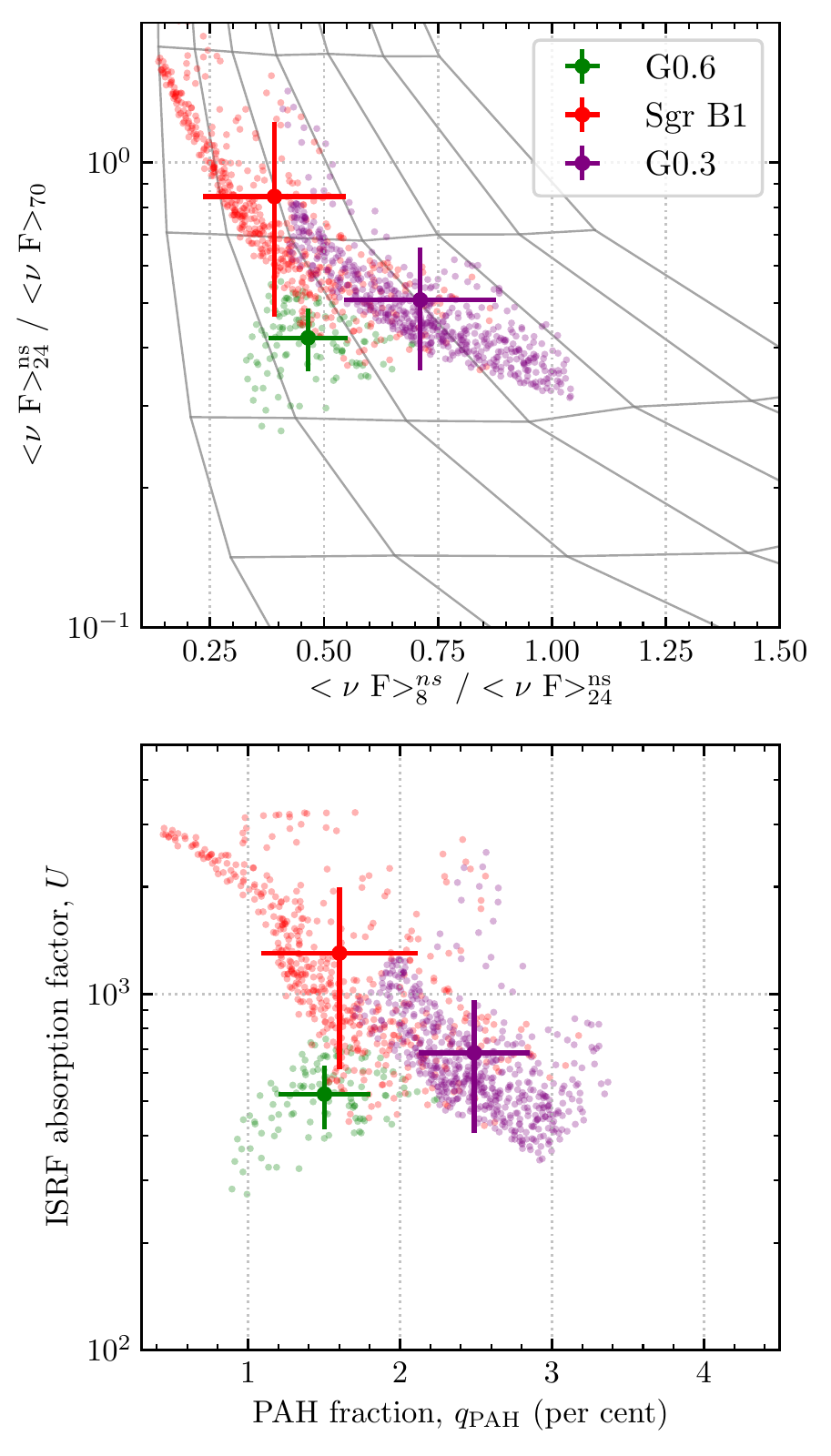}

\caption{Dust properties used to calculate the pressure component from the dust reprocessed emission. The upper panel shows the extinction and star-light contribution corrected infrared flux ratios for all pixels within each (proto)-cluster. We overplot a grid of the predicted flux density ratio from the dust model for various values of the fraction of dust in polycyclic aromatic hydrocarbons, $q_\mathrm{PAH}$, and the dimensionless scale factor of the radiation absorbed by the dust, $U$. Horizontal lines show $q_\mathrm{PAH}$ = 0.47, 1.12, 1.77, 2.50, 3.19, 3.90, 4.58 from left to right, and vertical lines show $U$ = 5, 15, 25, 100, 300, 1000, and 3000 from \note{bottom to top}. The lower panel shows the results from the interpolation of the $q_\mathrm{PAH}$ and $U$ dust model grid onto the observed flux ratios. The 5.8, 8 and 24\,\micron\ fluxes have been corrected for a constant visual extinction of $A_\mathrm{v}=20$\,mag \citep{figer_1999, dutra_2003, schodel_2010}. Sgr B2 has been omitted from this analysis due to the high infrared extinction ($A_\mathrm{8\micron}>10$\,mag, equal to a visual extinction of $A_\mathrm{v}>100$\,mag).}
\label{energy_denisty}
\end{figure}

The luminosity from the young massive stars peaks at ultraviolet wavelengths. However, this regime is completely obscured for most embedded star-forming regions. Rather, the ultraviolet emission from the majority of young stars is absorbed by the surrounding dust from the host molecular cloud. This then heats the dust from tens to hundreds of Kelvin, so that it emits predominantly at infrared wavelengths. The radiation field produced by the heated dust then provides an expansion pressure, which can be given as, 
\begin{equation}
P_\mathrm{IR} = \frac{1}{3} u,
\end{equation}
where $u$ is the radiation field absorbed by the dust. To estimate $u$, we compare the observed 5.8, 8, 24 and 70\,\micron\ \citep{churchwell_2009, carey_2009, molinari_2010} infrared flux densities to those predicted for the dust models of \citet{draine_2007}. These dust models produce synthetic spectral energy distributions for a range of radiation fields and dust compositions. The latter is parametrised by the fraction of dust in polycyclic aromatic hydrocarbons, $q_\mathrm{PAH}$, which produce substantial emission at the observed infrared wavelengths. The former is parameterised by the dimensionless scale factor of the radiation absorbed by the dust, $U$, which measures the radiation field energy density normalised to the local interstellar radiation field (ISRF) energy density,
\begin{equation}
u = U u^\mathrm{ISRF}.
\label{equation:energy density}
\end{equation}
Here, $u^\mathrm{ISRF}\,=\,8.65\,\times\,10^{-13}$\,erg\,cm$^{-3}$ is the energy density of non-ionising photons in the local interstellar medium.

To compare the observed 5.8, 8, 24 and 70\,\micron\ infrared flux densities for the \HII\ regions to those predicted by the dust models, we smooth the infrared maps to a common resolution of 11.5\arcsec. Following \citet{barnes_2017}, we then use the prescriptions of \citet{cardelli_1989} and \citet{chapman_2009} to correct the 5.8, 8 and 24\,\micron\ fluxes for a constant visual extinction of $A_\mathrm{v}=20$\,mag, which is typical for lines of sight towards the Galactic Centre \citep{figer_1999, dutra_2003, schodel_2010}. Additionally, we determine the extinction along each line of sight using the molecular hydrogen column density map (e.g. \citealp{molinari_2011}, Battersby et al. in prep). To do so, we use the conversion from column density to visual extinction from \citet{fitzpatrick_1999}, and then the conversions from visual to infrared extinction from \citet{cardelli_1989} and \citet{chapman_2009}. When doing so we find that Sgr B2 has a very high infrared extinction for the majority of its sight-lines ($A_\mathrm{8\micron}>10$\,mag, equal to a visual extinction of $A_\mathrm{v}>100$\,mag), which can not be accurately corrected. It has, therefore, been omitted from this analysis of the dust-processed radiation pressure. 

We remove the contribution of star-light from the 8 and 24\micron\ flux density maps using the 3.6\micron\ flux density map with \citep{lopez_2011, lopez_2014},
\begin{equation}
F_{8}^\mathrm{ns} = F_{8} - 0.232F_{3.6} 
\\
{\rm and}
\\
F_{24}^\mathrm{ns} = F_{24} - 0.032F_{3.6} 
\end{equation}
where the subscript denotes the wavelength, and superscript denotes the non-stellar flux densities. Due to the very small 3.6\micron\ flux density observed within the Galactic Centre, this subtraction has a minimal effect. 

Ratios of the observed 8, 24 and 70\micron\ fluxes ($F_{8}^\mathrm{ns}$, $F_{24}^\mathrm{ns}$ and $F_{70}$) are required to compare to the \citet{draine_2007} dust models. These ratios for all the pixels within each of the Galactic Centre \HII\ regions are shown in the upper panel of Figure\,\ref{energy_denisty}. We overplot on the measurements a grid of flux ratios predicted from the dust model. When corrected for a constant extinction, we find that G0.6, Sgr B1 and G0.3 have flux ratios well covered by the dust model grid. However, when we correct for the extinction along each line of sight using the column density measurements, several of the flux ratio values fall out of the dust model grid parameter space (e.g. to the right of the grid plotted in the upper panel of Figure\,\ref{energy_denisty}). We note that we are interested in the value of the radiation absorbed by the dust to calculate the dust processed radiation pressure, and the values of $U$ are close to constant for a given $F_{24}^\mathrm{ns}/F_{70}$ (i.e. horizontal grid lines). Therefore, we interpolate the $U$ value for the observed $F_{24}^\mathrm{ns}/F_{70}$ from the dust model grid for any values that fall out of the grid parameter space; i.e. ignoring any $F_{8}^\mathrm{ns}/F_{24}^\mathrm{ns}$ dependence.  

We use the {\sc interpolate.griddata}\footnote{http://docs.scipy.org} of {\sc scipy} package in {\sc ipython} to interpolate the grid of predicted flux ratios from the dust models onto the observed values. The lower panel of Figure\,\ref{energy_denisty} shows the values of $U$ as a function of $q_\mathrm{PAH}$ for all pixels from this interpolation process (for a constant extinction correction). The values of $U$ determined using a constant and varying extinction are used in equation\,\ref{equation:energy density} to determine the dust processed pressure component.

\note{It is worth briefly mentioning here that by using the dust model to get $q_\mathrm{PAH}$, we are effectively accounting for the high Galactic Centre metallicity when determining $U$ and, hence, $P_\mathrm{IR}$. The absolute metallicity or dust abundance does not matter very much for the \citet{draine_2007} models we are adopting. This is because the modelling we conduct here amounts to asking what spectrum the dust will emit, assuming it is exposed to a certain background radiation field and that the optical depth to the re-emitted infrared is small. Since only the spectral shape is being used in the calculation, doubling the number of grains per unit volume has no effect, since it does not change the spectral shape, just the absolute luminosity. There could be a subdominant effect in that the grain size distribution may be different for the Galactic Centre than it is at lower metallicity. This will make a difference, because the spectral energy distribution is sensitive to the grain size distribution. However, there is no evidence that the grain size distribution is significantly different for super-solar metallicity, and, to the extent that it is, the first order-variation is captured by our variation of the PAH fraction ($q_\mathrm{PAH}$).}

Again, as in section\,\ref{subsec:Direct radiation}, we adopt the two methods for calculating the pressure as a function of the size scale. Firstly, we take the \HII\ region (i.e. leaf boundary) masks produced from the dendrogram analysis, and apply these to the $U$ maps for each source to get a measure $P_\mathrm{IR}$ within distinct \HII\ regions of different sizes. Secondly, we calculate the average $U$ within circular apertures of increasing radius, which gives a radial dependence of $P_\mathrm{IR}$ from the centre of each source. 

\begin{figure*}
\centering
\includegraphics[trim = 0mm 0mm 0mm 0mm, clip,angle=0,width=1.0\textwidth]{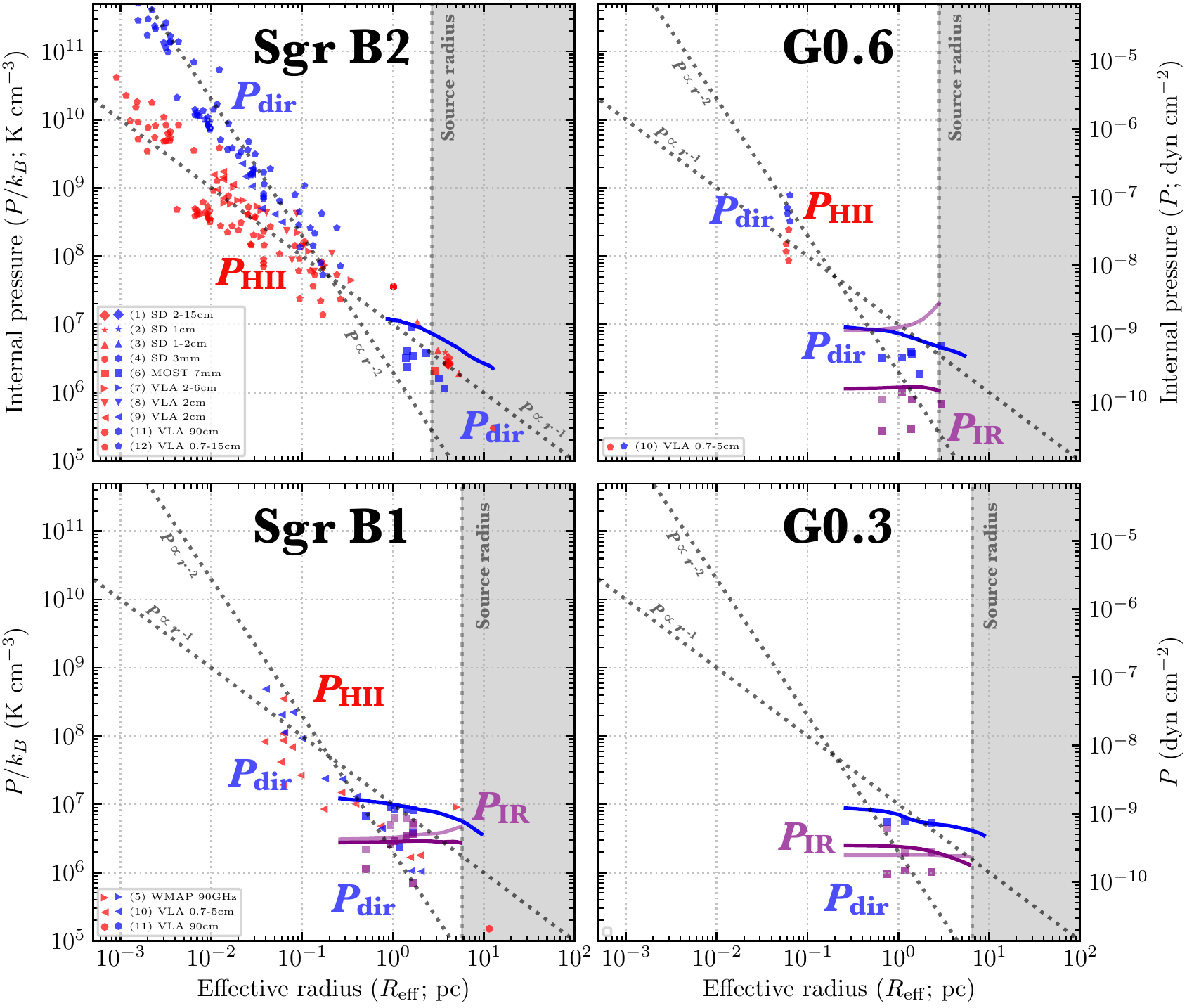}
\caption{The pressure as a function of effective radius for each source. \note{The points within this plot represent discrete measurements of the pressure calculated from sources identified from various resolution radio data sets that have been taken from the literature (see Table\,\ref{radio_obs}), or from the sources identified within the available infrared observations (see sections\,\ref{subsec:Direct radiation} and \ref{subsec:Dust reprocessed emission}). The lines represent radial profiles from the centre of each source, which have been determined using the infrared observations (sections\,\ref{subsec:Direct radiation} and \ref{subsec:Dust reprocessed emission}).} Blue points and lines show the direct radiation pressure ($P_\mathrm{dir}$; section\,\ref{subsec:Direct radiation}), purple points and lines show the dust reprocessed emission pressure ($P_\mathrm{IR}$; section\,\ref{subsec:Dust reprocessed emission}), red points indicate the warm ionised gas pressure ($P_\mathrm{HII}$; section\,\ref{subsec:Warm ionised gas}). The faded purple lines and points show the result of using the extinction determined along each line of sight, as opposed to using a constant value, for the $P_\mathrm{IR}$ analysis (see section\,\ref{subsec:Dust reprocessed emission}). The legend shows the references for the radio data used to determine these pressure components, where the numbers correspond to the reference IDs provided in Table\,\ref{radio_obs}. We overlaid diagonal lines corresponding to $P\,\propto\,r^{-1}$ and $P\,\propto\,r^{-2}$ for reference. The vertical black dashed line is the measured effective radius for each source (see Table\,\ref{source_properties}). Points within the shaded grey region are larger than the measured source sizes, and, therefore, may have spatial overlap with adjacent sources (see source ellipses shown in Figure\,\ref{three_col_map}). Analysis within the shaded portion of the parameter space should be treated with caution. \note{The discrete measurements of each pressure component, shown as  points within this plot, are given in Table\,\ref{tab:pressurecomps} and a machine-readable table in the online supplementary material of this work.}}
\label{pressure_vs_radius}
\end{figure*}

\subsection{Total internal HII region pressure}\label{subsec_totalpressure}

Using the methods presented in this section, we have calculated the ionised gas ($P_\mathrm{HII}$), direct radiation ($P_\mathrm{dir}$), and dust reprocessed ($P_\mathrm{IR}$) pressure components for our sample of Galactic Centre \ion{H}{II} regions. Figure\,\ref{pressure_vs_radius} shows how these various pressure components vary as a function of the size scale for each of the Galactic Centre sources. The direct radiation pressure is shown by the blue lines and points, the warm ionised gas pressure is shown by the red points, and the dust reprocessed emission pressure is shown by the purple lines and points. The numbers within the legend correspond to the references provided within Table\,\ref{radio_obs}. 

On the smallest scales ($\sim10^{-3}$\,pc), the direct radiation pressure is factors of several higher than the ionised gas thermal pressure. However, we find that the direct radiation pressure decreases steeply with increasing radius (see dotted lines overlaid on each panel), and on scales of $\sim10^{-2}$\,pc both the direct and ionised gas pressure components are equal within the observed scatter. On intermediate scales (0.01-0.1pc; Sg B2, G0.6, Sgr B1), we find that the direct radiation and ionised thermal pressure components remain approximately equal within around an order of magnitude scatter. On the large scale (> 0.1-20pc), we find that the direct radiation is typically less than both the ionised thermal pressure gas and the dust-processed pressures, which are comparable. Where measurements are available for all three of the pressure components on the largest scales (G0.6 and Sgr B1), we find that the ionised thermal pressure component is the dominant pressure term driving the \HII\ region expansion.

It is worth considering at this stage how the radial trends displayed in Figure\,\ref{pressure_vs_radius} should be interpreted. Firstly, the $P_\mathrm{dir}$ and $P_\mathrm{HII}$ points have been calculated in two ways: one  method using radio observations, and another method using IR observations. The former provides measurements of the same \HII\ regions at different spatial resolutions. The latter uses dendrograms to identify distinct \HII\ regions, each of which can have a different physical size depending on the dendrogram analysis. Therefore, increasing in radius in Figure\,\ref{pressure_vs_radius} can mean that either (i) distinct increasingly larger \HII\ regions have been identified within each source, or (ii) these larger radii data points are larger aperture averages of the smaller effective radii \HII\ regions that they contain. The fact that the radial trend is the same for both methods using two different, widely separated wavelength observations show the results are robust against the choice of method. We conclude that a simplistic but reasonable interpretation of the points plotted within Figure\,\ref{pressure_vs_radius} are that they represent pressure components as a function of \HII\ region size.  

This is subtly different from the lines plotted on Figure\,\ref{pressure_vs_radius}, which have been determined from increasing size apertures from the centre of each \HII\ region. The lines then represent a radial dependence of the pressure components. In the interest of a comparison, both of the methods have been plotted on Figure\,\ref{pressure_vs_radius}, yet caution should be taken in drawing conclusions based on both the radial (lines) and size (points) distributions. 
\section{Pressure profiles and feedback-driven dynamics}\label{sec_analysis}

\subsection{Pressure components as radially decreasing power laws}\label{subsec_fitpressure}

\begin{table}
\centering
\caption{The pressure components as a function of radius power-law fit parameters as shown in Figure\,\ref{pressure_vs_radius_idv}. These have been determined using a least-squares fitting routine in log-log space for $P = a (R_\mathrm{eff}/\mathrm{pc})^{-b}$, where $a$ and $b$ are tabulated as the constant and power in this table (see section\,\ref{subsec_fitpressure}). This analysis has been limited to the warm ionised gas pressure and direct radiation pressure, and for sources that have pressure components determined over a sufficient range in effective radius to allow for statistically significant power-law fits.}
\begin{tabular}{c c c c c c c c c c c c c c c c c c c c}
\hline
Pressure & Sources & Power; $b$ & Constant; log($a$) \\
& & & (K\,cm$^{-3}$) \\
\hline
$P_\mathrm{HII}$ & Sgr B2 & -0.99$\pm$0.03 & 7.02$\pm$0.06 \\
$P_\mathrm{HII}$ & Sgr B1 & -0.94$\pm$0.14 & 6.64$\pm$0.13 \smallskip \\
$P_\mathrm{dir}$ & Sgr B2 & -1.68$\pm$0.04 & 6.79$\pm$0.07 \\
$P_\mathrm{dir}$ & G0.6 & -1.51$\pm$0.15 & 6.78$\pm$0.12 \\
$P_\mathrm{dir}$ & Sgr B1 & -1.27$\pm$0.1 & 6.71$\pm$0.08 \\
\hline
\end{tabular}
\label{fit_props}
\end{table}



\begin{figure}
\centering
\includegraphics[trim = 0mm 0mm 0mm 0mm, clip,angle=0,width=1.0\columnwidth]{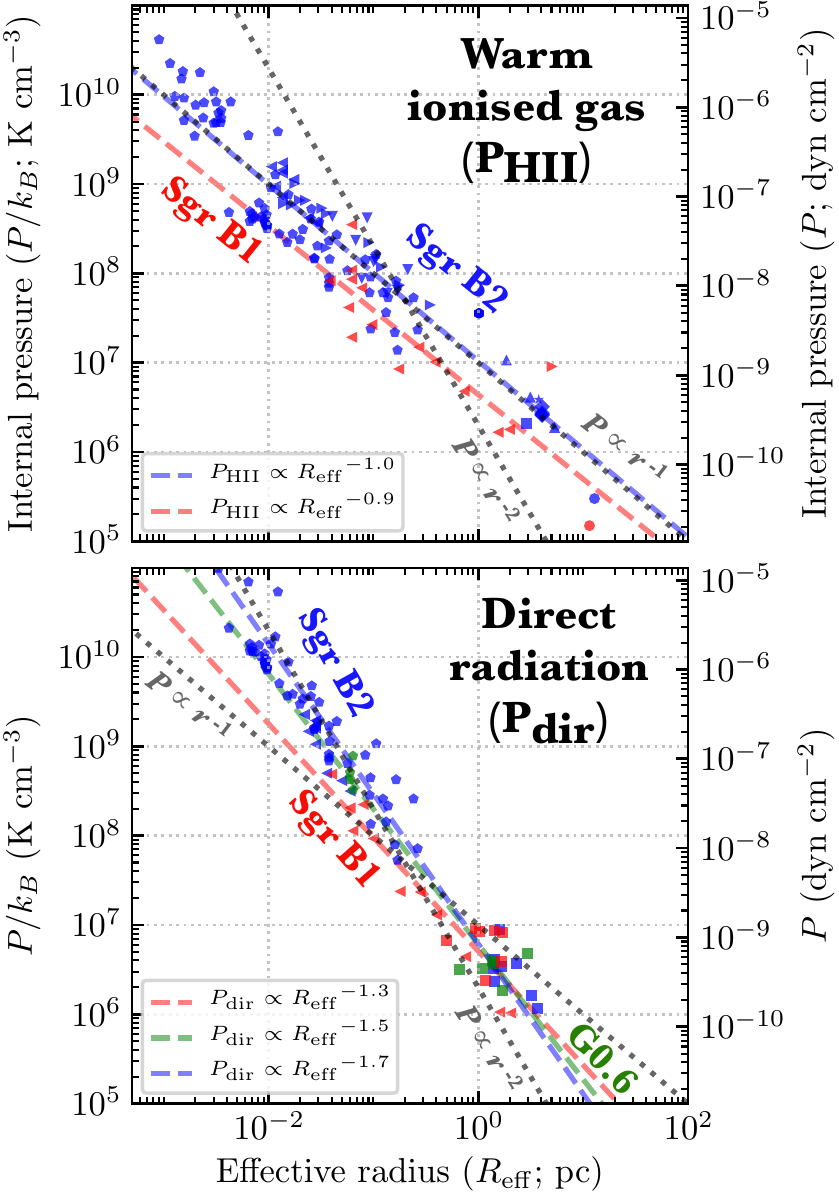}
\caption{The observed pressure as a function of radius for each source (labelled at the top of each panel). The symbols for each observed component are identical to those shown in Figure\,\ref{pressure_vs_radius}. Shown as diagonal dashed lines are the power-law fits to the pressure components for each source. The parameters of these fits are given in the legend of each panel, and are summarised in Table\,\ref{fit_props}. The diagonal grey dotten line shows $P\,\propto\,r^{-1}$ and $P\,\propto\,r^{-2}$ for reference.}
\label{pressure_vs_radius_idv}
\end{figure}

To examine how the pressure components determined in section\,\ref{sec_results} scale with effective radius (size), we perform a power-law fit to the points shown in Figure\,\ref{pressure_vs_radius}. We fit a power-law relation of $P = a (R_\mathrm{eff}/\mathrm{pc})^{-b}$ to each of the pressure components, using the {\sc numpy.polyfit} least-squares fitting routine in log-log space. This analysis \note{is} limited to the warm ionised gas pressure and direct radiation pressure, and for sources which \note{have} pressure components determined over a sufficient range in effective radius to allow for statistically significant power-law fits. Figure\,\ref{pressure_vs_radius_idv} displays the observed pressure components overlaid with the results of the fitting routine (as labelled). The parameters of each fit are given in Table\,\ref{fit_props}.

We find that the thermal pressure of the ionised gas ($P_\mathrm{HII}$) within the Sgr B2 and Sgr B1 sources are both best fit with the relation $P_\mathrm{HII}\propto r^{-1}$. We compare this to what is expected from simple analytic arguments. The ideal gas equation is given as $P_\mathrm{therm}\propto n T$, where $n$ is the number density, and $T$ is the temperature. 
The Stromgren radius of an \HII\ region can be expressed as $r^3\propto \mathcal{N}_\mathrm{LyC} n^{-2}$, where $\mathcal{N}_\mathrm{LyC}$ is the flux of ionising photons (see equation\,\ref{stromgren_R}; \citealp{stromgren_1939}). Therefore, we would expect $P_\mathrm{therm}\propto r^{-3/2}$. We then find that the observed ionised gas pressure decreases less steeply with radius than estimated from these simple thermal pressure arguments (see section\,\ref{simulations} for further discussion).  

We find the direct radiation pressure has a significantly steeper radial trend ($P_\mathrm{dir} \propto R_\mathrm{eff}^{-1.5}$) compared to the ionised gas pressure ($P_\mathrm{HII}\propto R_\mathrm{eff}^{-1}$). Moreover, we find that there is moderate ($\pm$0.2 in the exponent of the powerlaw slopes) source-to-source variation for the direct radiation pressure radial relations, which is not observed for the ionised gas pressure. Comparing to this variation to the ages of the \HII\ regions, we infer that the slope of the direct radiation pressure becomes shallower with age (compare Tables\,\ref{source_properties} and \,\ref{fit_props}). 

\subsection{Determining the expansion velocity, momentum and energy of the \ion{H}{II} regions}\label{sec:energetics}

We now want to measure the expansion velocity of the \HII\ regions and the associated energy and momentum required to drive the \HII\ region bubbles.

To obtain an accurate expansion velocity for the star-forming clouds, we measure both a photometric expansion rate from the {\it Spitzer} and {\it Herschel} MIR luminosity maps, and a spectral expansion rate from radio recombination lines (RRLs).\footnote{Here we use the standard RRL notation of H$n_{\upalpha}$, which corresponds to a downward energy level transitions from principal quantum $n+1$ to $n$ for hydrogen (e.g. $n=54\rightarrow53$ for hydrogen is H53$_{\upalpha}$)}. These measurements should represent expansion velocities perpendicular to the line-of-sight and along the line-of-sight, respectively. 

\begin{table}
\centering
\caption{Photometrically (section\,\ref{sec_photo}) and spectroscopically (section\,\ref{sec_spec}) determined expansion velocities ($v_\mathrm{exp}$) for the Galactic Centre \HII\ regions. Also tabulated are the (log) energies, and (log) momenta estimated using the average expansion velocities and mass of each \HII\ region (section\,\ref{sec_ep}).}
\begin{tabular}{c c c c c}
\hline
Source & $v_\mathrm{exp,pho}$ & $v_\mathrm{exp,spec}$ & log($E_\mathrm{exp}$) & log($p_\mathrm{exp}$) \\ 
& (km\,s$^{-1}$) & (km\,s$^{-1}$) & (J) & (\sol\,km\,s$^{-1}$) \\ 
\hline



Sgr B2 & 12.8 & 10.7 & 41.75 & 4.69 \\
G0.6 & 3.0 & - & 41.48 & 5.00 \\
Sgr B1 & 5.6 & 9.2 & 42.20 & 5.33 \\
G0.3 & 5.2 & - & 41.92 & 5.21 \\

\hline
\end{tabular}
\label{props_ep}
\end{table}

\subsubsection{Photometric velocity}\label{sec_photo}

The model of \citet{kruijssen_2015} predicts that the clouds and \HII\ regions with the CMZ may reside along an orbital stream. In this model, star formations began at pericentre passage with the bottom of the Galactic gravitational potential, when compressive tidal forces are strongest \citep{longmore_2013a, kruijssen19b, dale19}. On this orbit is the dust-ridge (shown in Figure\,\ref{three_col_map}), which shows increasing signs of active star formation from the point of pericentre passage. Cloud e/f (see Figure\,\ref{three_col_map}) is the first cloud in this sequence that shows several signs that star formation has recently begun (e.g. maser emission) so we use this as the zero point for star formation activity. We use the difference in time along the orbit from cloud e/f to each of the ionised ridge sources as an estimate of the expansion time of the \HII\ regions. Referring to the nomenclature in \citet{kruijssen_2015}, the age of the \HII\ region is given by $t_\mathrm{age} = t_\mathrm{p,last} - t_\mathrm{p,last}(\mathrm{cloud\,e/f})$. These ages are given in Table\,\ref{source_properties}, and displayed in the inset of Figure\,\ref{three_col_map}. With these, we can give a photometric estimate of the expanding shell velocity, which is defined as,
\begin{equation}
v_\mathrm{exp, pho} = \frac{R_\mathrm{eff}}{t _\mathrm{age}},
\end{equation} 
The results of this analysis are displayed in Table\,\ref{props_ep}.

\subsubsection{Spectral velocity}\label{sec_spec}

\begin{figure*}
\centering
\includegraphics[trim = 0mm 0mm 0mm 0mm, clip,angle=0, width=1\textwidth]{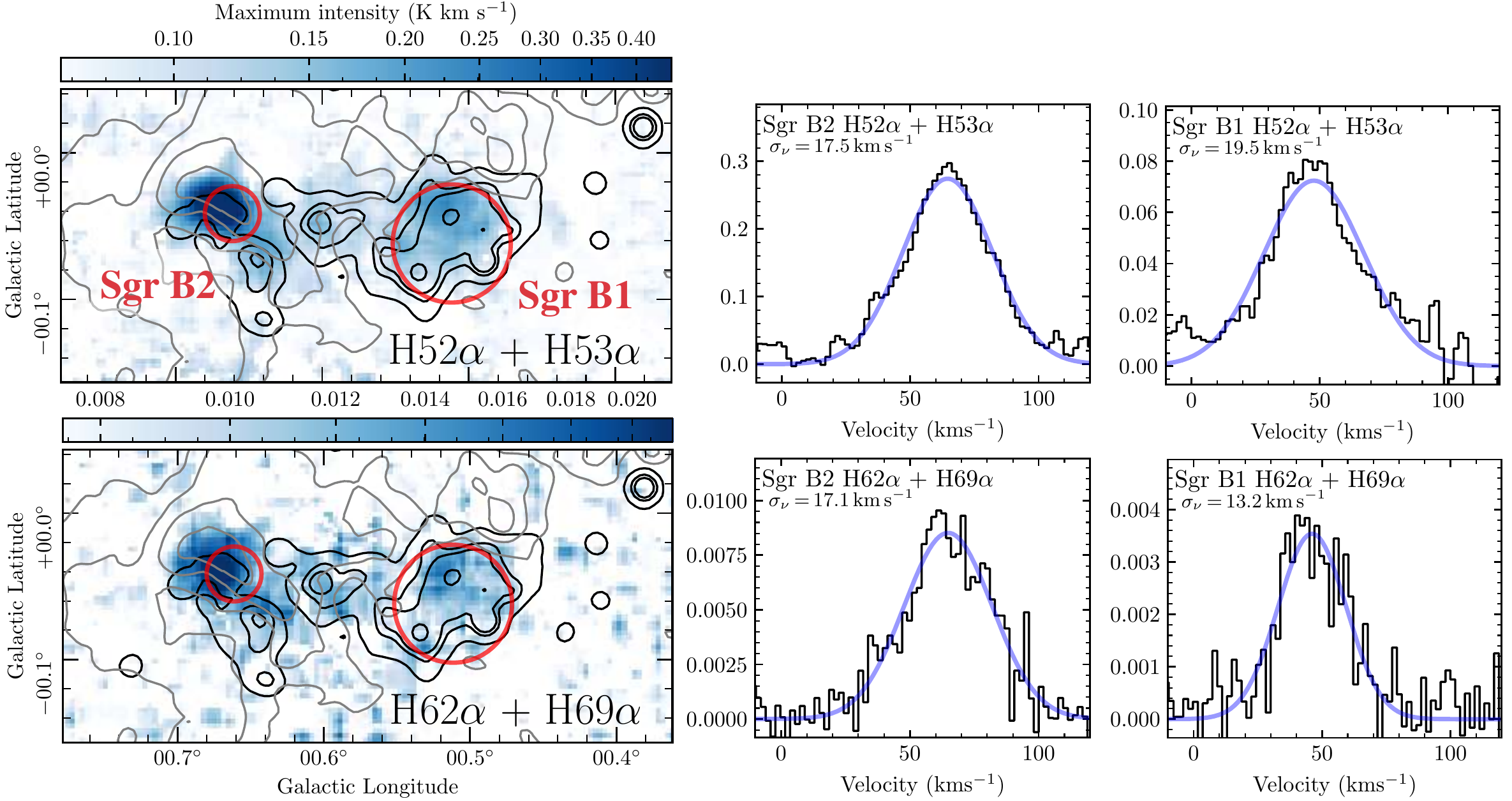}
\caption{[Left panels] The radio recombination lines that have been used to investigate the \ion{H}{II} region expansion velocities. These have been taken from the Mopra CMZ survey \citep{walsh_2008, walsh_2011, purcell_2012, jones_2013}, and have been smoothed to the same spatial and spectral resolution. To achieve better signal-to-noise, the RRLs with similar frequencies have been stacked (H52$\alpha$\,+\,H53$\alpha$; H61\,+\,H69$\alpha$). [right panels] Mean spectra across Sgr B2 and Sgr B1 \citep{barnes_2017}. These have been fit with a single Gaussian velocity component, which is shown as a faded blue curve on each spectrum, and the velocity dispersion from each fit is shown at the top of each panel.} 
\label{spec}
\end{figure*}

As an independent measure of the expansion velocity along the line-of-sight, we can also use spectroscopic observations. There have been many surveys that have observed the molecular line emission and studied the dynamics of the cool gas within the Galactic Centre. However, studies of the \ion{H}{II} region dynamics, and how these fit into the current scenarios of Galactic Centre kinematics are currently lacking. To investigate the ionised gas within the \ion{H}{II} regions, we use H52$\alpha$, H53$\alpha$,  H61$\alpha$, and H69$\alpha$ recombination line observations taken with the Mopra telescope \citep{walsh_2008, walsh_2011, purcell_2012, jones_2013}. Initial inspection of these observations \note{show} that the signal-to-noise ratio towards the \ion{H}{II} regions \note{is} not sufficient to accurately determine kinematic information. To overcome this, we stack the radio recombination lines from the survey that \note{are} close in frequency, and then from these stacked maps \note{choose} those which \note{have} good signal-to-noise values. 

The maps towards the ionised ridge that have been integrated between $0 - 100$\,\kms\ are shown in the left-hand panel of Figure\,\ref{spec}. These clearly show a strong detection towards the Sgr B2 and Sgr B1 sources, yet lack emission towards G0.6 and G0.3. We, therefore, only analyse Sgr B2 and Sgr B1 in this section. The spectra taken towards the Sgr B2 and Sgr B1 sources from each of these stacked maps are shown in the right-hand panel of Figure\,\ref{spec}. These stacked spectra have been fit with a Gaussian profile using the {\sc pyspeckit} package in {\sc python}. This fit is overlaid on each of the spectra shown in Figure\,\ref{spec}, and the measured velocity dispersion, $\sigma_\mathrm{v}$, is shown in the upper left of each panel.  

To calculate the expansion velocity of each of the \ion{H}{II} regions from their measured velocity dispersion, $\sigma_\mathrm{v}$, we assume that there are several distinct contributions that combined in quadrature to give the observed Gaussian profile. The first of these is the thermal motions of the ionised gas, $\sigma_\mathrm{T}\sim\,\sqrt{2.2\,T_\mathrm{e}k_\mathrm{B}/m_\mathrm{H} \mu_\mathrm{H}}\,=\,8.05$\,\kms, \note{which assumes a mean atomic weight accounting for helium ($\mu_\mathrm{H}=1.41$), a factor of 2.2 that accounts for the number of particles (including electrons) per H nucleus, and the typical Galactic Centre \ion{H}{II} region electron temperature of 5000\,K} (see section\,\ref{subsec:Warm ionised gas}). The second of these contributions is the dynamical motions (e.g. rotation, relative motions) inherited from earlier in their evolution as molecular clouds: $\sigma_\mathrm{D}$. This is taken as the average dispersion of the ``dust-ridge'' of 10.8\,\kms\ \citep{henshaw_2016}. Lastly, we subtract the contribution from the velocity resolution $v_\mathrm{res} \sim 0.2$\,\kms. Finally, following \citet[][their equation\,3]{keto_2008}, we determine the contribution from pressure broadening, $\sigma_\mathrm{P}$, assuming a $n_\mathrm{e}=10^{3}$\,cm$^{-3}$, which corresponds to the electron density calculated for around a parsec scale within Sgr B2 and B1 (using equation\,\ref{mh} and the flux density measurement of \citealp{downes_1970} and \citealp{mehringer_1992}). We find that the contribution of $\sigma_\mathrm{P}$ to $\sigma_\mathrm{v}$ is $<1$\,per cent for the studied radio recombination lines, and, therefore, $\sigma_\mathrm{P}$ not considered further. The expansion velocity, $v_\mathrm{exp,spec}$, for each of the \ion{H}{II} regions is then calculated as,
\begin{equation}
\sigma_\mathrm{v}\,=\,\sqrt{\sigma_\mathrm{D}^2\, + \,\sigma_\mathrm{T}^2\, + \,\Delta v_\mathrm{res}^2 \, + \,v_\mathrm{exp,spec}^2},
\label{eq:photo_expansion} 
\end{equation} 
The results of this analysis are given in Table\,\ref{props_ep}. On average we find spectroscopic expansion velocities for Sgr B2 and B1 of the order 10\,\kms, \note{and both fall within around 50~per~cent of the photometric velocities. Given the assumptions made in the derivations of these values, this represents satisfactory agreement.}

\subsubsection{Energy and momentum}\label{sec_ep}

In this section, we have determined the expansion velocity of the \ion{H}{II} regions using two independent methods, which \note{gives} values of $v_\mathrm{exp}$ for each of the sources within reasonable agreement. The average of these expansion velocities can now be used to calculate how much energy and momentum from the embedded stellar population is imparted on the surrounding environment. 

The total energy, $E_\mathrm{exp}$, and momentum, $p_\mathrm{exp}$, of the expanding gas can be simply calculated as,
\begin{equation}
\begin{split}
&E_\mathrm{exp} = \frac{1}{2} \, M_\mathrm{ejct} \, v_\mathrm{exp}^{2} \\
&p_\mathrm{exp} = M_\mathrm{ejct} \, v_\mathrm{exp},
\end{split}
\end{equation}
where $M_\mathrm{ejct}$ is, 
\begin{equation}
M_\mathrm{ejct} = M^\mathrm{init}_\mathrm{gas} - M^\mathrm{final}_\mathrm{*},
\label{equ_ejct_mass}
\end{equation}
where $M^\mathrm{init}_\mathrm{gas}$ is the gas mass of the precursor molecular clouds from which the \note{stellar population} formed, and $M^\mathrm{final}_\mathrm{*}$ is the final stellar mass of the \note{stellar population}. Given gas clouds with properties similar to those currently in the dust ridge are the most likely precursors clouds to the \note{stellar populations} driving the \HII\ regions, we use the average mass of the dust ridge clouds as our estimate of $M^\mathrm{init}_\mathrm{gas}$ (see Table\,\ref{source_properties}).

Here we have implicitly made the assumption that the ejecta mass can be represented by the difference in mass between the average dust ridge cloud and the embedded stellar mass within an \ion{H}{II} region. Or in other words, all the initial cloud mass that is not converted to stars, is blown away by the \ion{H}{II} region as ejecta. We note that, however, this scenario will not strictly be the case, given that we know there is still some dust continuum emission towards the \ion{H}{II} regions. This is particularly relevant for Sgr B2, which is thought to be still heavily embedded within its host cloud (see Figure\,\ref{three_col_map}). In light of this, the values of ejecta mass, energy and momentum calculated in this section should be viewed as upper limits. The resulting energy and momentum values are shown in Table\,\ref{props_ep}. We discuss these values in section\,\ref{energy}.
\section{Discussion}\label{sec_discussion}

\subsection{Comparison with lower pressure environments}\label{sec:comp2SMCLMC}

\begin{figure}
\centering
\includegraphics[trim = 0mm 0mm 0mm 0mm, clip,angle=0,width=1.0\columnwidth]{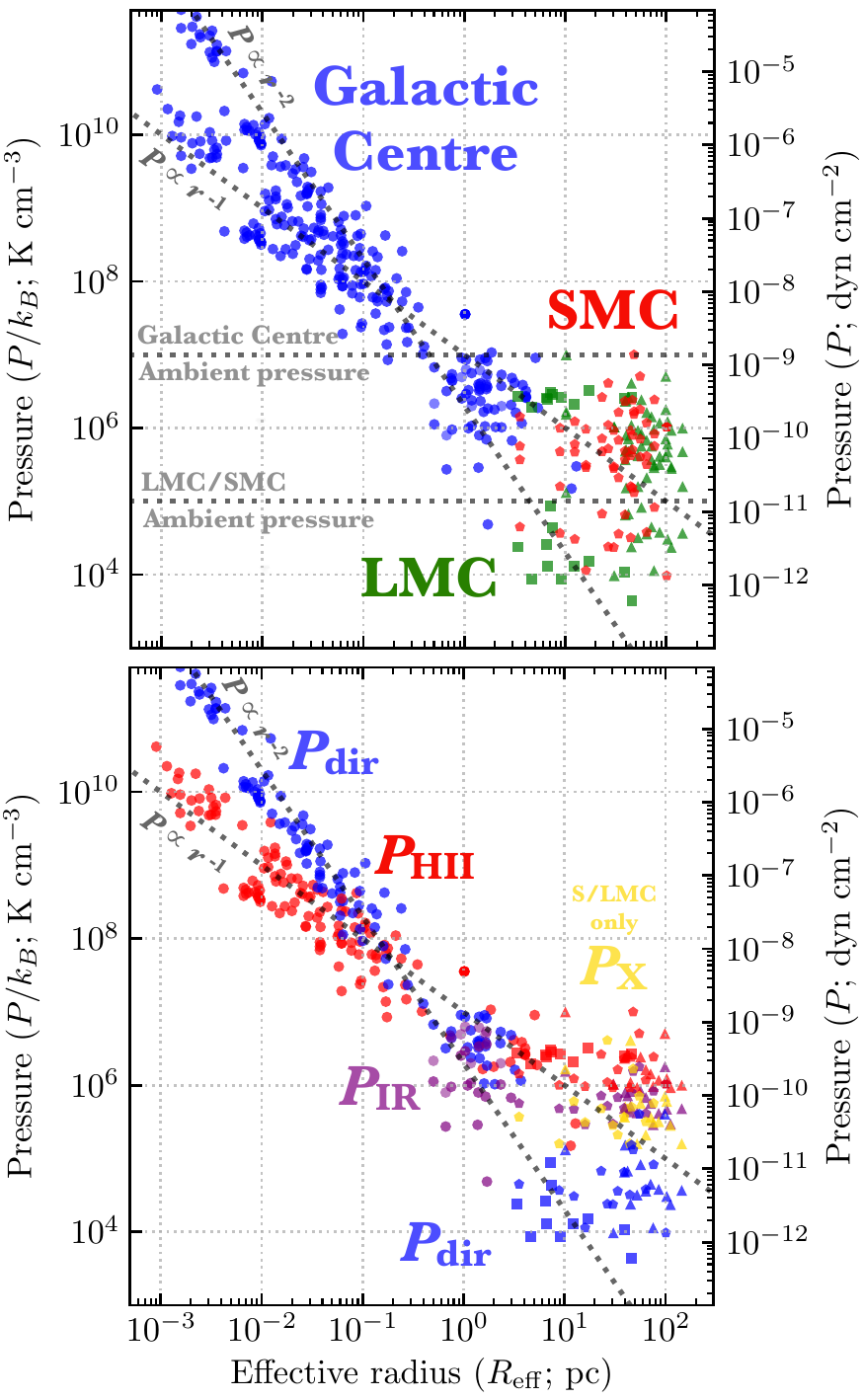}
\caption{A comparison between the Galactic Centre, LMC and SMC \citep{lopez_2011, lopez_2014, mcleod_2019}. In the upper panel the points are differentiated in colour by their host environment, and in the lower panel the points are differentiated by the pressure component (as labelled). Circles show the pressure components determined in this work, pentagons are SMC sources from \citet{lopez_2014}, triangles are LMC sources from \citet{lopez_2011, lopez_2014} and squares are LMC sources from \citet{mcleod_2019}. \note{We note that the hot X-ray emitting gas pressure component ($P_\mathrm{X}$) was not calculated within the CMZ due to the high extinction (section\,\ref{sec:Pressure}), but we do show $P_\mathrm{X}$ determined for the LMC and SMC for comparison \citep{lopez_2011, lopez_2014}.} We show the ambient pressures determined for the Galactic Centre \citep{walker_2018} and LMC \citep{hughes_2013} as horizontal grey dotted lines. The diagonal grey dotted lines show the $P\,\propto\,r^{-1}$ and $P\,\propto\,r^{-2}$ relations.}
\label{fig_MCcomp}
\end{figure}

We now compare the pressure components calculated for the Galactic Centre \HII\ regions to those found in other similar observations in the literature. Currently, however, \note{\HII\ regions within a limited sample of sources have been investigated in a comparable manner (e.g. \citealp{lopez_2011, lopez_2014, mcleod_2019, mcleod_20}).} We make use of the data taken from, \citet{lopez_2011} who investigated the pressure components within the massive star-forming region 30 Doradus in the Large Magellanic Cloud, and \citet[][data taken from their Table\,7]{lopez_2014} who then expanded this study to 32 \HII\ regions with ages of $\sim\,3-10$\,Myr within both the Small and Large Magellanic Clouds (SMC and LMC, respectively). More recently, \citet[][data are taken from their Table\,7]{mcleod_2019} have used MUSE integral field data to accurately determine the spectral types and luminosity classes of the stellar populations within two \HII\ region complexes in the LMC.\footnote{We note that the direct radiation pressure calculated by \citet{mcleod_2019} has been increased by a factor of three to match the definition outlined by \citet{lopez_2011, lopez_2014}, and this work (see equation\,\ref{eq_Pdir}).} From this they determine the direct radiation and ionised gas pressure components of the feedback. The upper panel of Figure\,\ref{fig_MCcomp} shows how the various pressure components for \HII\ regions within the LMC and SMC (green and red points, respectively) compare to those within the Milky Way Galactic Centre (blue points). 

The first thing to note about the comparison shown in Figure\,\ref{fig_MCcomp} is that the \HII\ regions within the LMC and SMC have systematically larger effective radii and lower pressures than the Galactic Centre \HII\ regions. We find that the Galactic Centre \HII\ regions extend to an effective radius of several parsecs (average of 4.5\,pc from Table\,\ref{source_properties}, while the sources within the SMC/LMC have average radii of around $\sim$\,50\,pc, with several sources within the LMC extending out to radii well above 100\,pc; comparable to the size of the entire Galactic Centre (see Figure\,\ref{three_col_map}). 

It is interesting to compare these maximum sizes to the scales when the ambient pressure equals the observed internal pressures. \citet{walker_2018} estimate molecular clouds within the Galactic Centre are subjected to an external pressure of the order $P/k_\mathrm{B}$\,$\sim$\,$10^{7-8}$\,K\,cm$^{-3}$, whilst the external pressures found within LMC/SMC and Milky Way disc environments is typically \note{at least} two-to-three orders of magnitude lower ($P/k_\mathrm{B}$\,$\sim$\,$10^{5}$\,K\,cm$^{-3}$; \citealp{bertoldi_1992, lada_2008, field_2011, belloche_2011, hughes_2013}). These ambient pressures are overlaid on Figure\,\ref{fig_MCcomp} as horizontal grey dotted lines. We find that the ambient pressure for the Galactic Centre crosses the $P\,\propto\,r^{-1}$ relation at a radius of several parsecs, whilst the LMC/SMC ambient pressure crosses this relation at a radius of few hundred parsecs. These values are then broadly comparable to the observed maximum sizes of the \HII\ regions within the two environments. This would suggest that the maximum size of the \HII\ regions is set by the point where the internal and ambient pressure are equal, and hence the higher ambient pressure within the Galactic Centre is limiting the \HII\ regions to a smaller size. The trend we observe cannot be due to the low angular resolution of the LMC/SMC observations. For example, 30 Doradus is a well studied \HII\ region complex within the LMC, and has a resolved radius of $\sim$\,200\,pc \citep{lopez_2011}, an order of magnitude larger than any resolved \HII\ regions within the Galactic Centre. However, we note that while this lack of large \HII\ regions in the Galactic Centre is likely physical, the absence of small ones from the LMC/SMC samples is probably affected by resolution limits. The ATCA map used in \citet{lopez_2014} has a resolution of 22\arcsec\,$\sim$\,5\,pc at the distance of the LMC, hence the Galactic Centre \HII\ regions would be confined to a single pixel in their LMC map.

The lower panel of Figure\,\ref{fig_MCcomp} shows how the various pressure components compare irrespective of the host environment. We find that the pressure components broadly follow the previously discussed radial dependences (section\,\ref{subsec_totalpressure}). We see that the direct radiation pressure follows a steep slope, from being the dominant pressure term at scales of $<0.01$\,pc to becoming almost negligible at $\gg1$\,pc scales. The remaining pressure terms appear to be approximately equal within an order of magnitude scatter, and show no appreciative decline as a function of the radius at $>10$\,pc scales.  

\subsection{Comparison to expansion rates from analytic models}\label{sec:comp2analmodels}

Many analytic models in the literature estimate the expansion rate of \HII\ regions (e.g. \citealp{bisbas_2015, geen_2019}). In this section, we investigate how well the sizes of the Galactic Centre \HII\ regions are predicted by several of the commonly used expansion models given assumptions about the age of each \HII\ region. In doing so, we aim to test the importance of including the various pressure components into these analytic models. For brevity, we only introduce the final analytic solutions of each model below, and refer the interested reader to the original papers for full details. 

The simplest and mostly widely known of the expansion models that takes into account only the thermal pressure of the warm ionised gas was outlined by \citet{spitzer_1978}, and later modified by \citet{dyson_1980}. This model assumes an initially spherically symmetric cloud with a radius of $R_\mathrm{cl}$, a total mass of $M_\mathrm{cl}$ and a uniform density of $\rho_\mathrm{cl}$ containing atomic hydrogen with a uniform temperature of $T_\mathrm{cl}$. As shown by \citet{stromgren_1939}, the flux of high energy Lyman continuum photons, $\mathcal{N}_\mathrm{LyC}$, emitted by the central source will ionise a spherical region of radius,
\begin{equation}
R_\mathrm{st} = \left(\frac{3 \mathcal{N}_\mathrm{LyC} m^{2}_\mathrm{p}}{4 \pi \alpha_\mathrm{B} \rho_\mathrm{cl}^{2}} \right)^{1/3} = 0.57 \left( \frac{\mathcal{N}_\mathrm{LyC}}{10^{50}\mathrm{s^{-1}}} \right)^{1/3} \left( \frac{\rho_\mathrm{cl}}{100\mathrm{M_\odot pc^{-3}}} \right)^{-2/3},
\label{stromgren_R}
\end{equation}
where $m_\mathrm{p}$ is the mass of a proton, and the recombination coefficient can be taken as $\alpha_\mathrm{B} \approx 2.7\times10^{-13}$ cm$^3$ s$^{-1}$. This simplified solution for $R_\mathrm{st}$ assumes an ISM that is composed solely of H, with no He. The so-called Spitzer solution to the variation in \HII\ region radius with time is given as, 
\begin{equation}
R_\mathrm{Sp}(t) = R_\mathrm{st} \left( 1 + \frac{7}{4} \frac{\sigma_\mathrm{T,i} t}{R_\mathrm{st}} \right) ^{4/7}
\label{spitzer}
\end{equation}
where $\sigma_\mathrm{T,i} $ is the sound speed within the ionised gas \note{($\sim\,8$\,\kms\ at 5000\,K; see section\,\ref{subsec:Warm ionised gas})}, and $t$ the age of the \ion{H}{II} region. \note{For all analytic solutions presented in this section, we make the assumption that $T_\mathrm{cl}/T_\mathrm{HII}\ll$1, where the $T_\mathrm{cl}$ is the neutral medium temperature and $T_\mathrm{HII}$ is the ionised gas temperature. Nonetheless, we consider that this may not hold within the Galactic Centre, where both the electron temperature is lower and the neutral gas temperature is higher; $T_\mathrm{cl}/T_\mathrm{HII}$\,$\sim$\,50\,K/5000\,K\,$\sim$\,0.01 (e.g. \citealp{mehringer_1992, depree_1996, clark_2013, krieger_2017}). To test this, we use the integral of the full Spitzer solution given in \citet[][equ. 9 and 10]{raga_2012}, and find an overall decrease in the predicted size of the \HII\ regions of 1 and 10\,per cent at times of 0.2 Myr and 1.3 Myr, respectively. This deviation is well within the expected uncertainty on these analytic solutions inherited from the broad range of input parameters, and, therefore, continue to work under the assumption that the $T_\mathrm{cl}/T_\mathrm{HII}\ll$1 simplification holds within the Galactic Centre.}

A second simple method for describing the expansion of an \ion{H}{II} region is provided by \citet{Hosokawa_2006}, who used the equation of motion of the expanding shell to derive the time dependent position of the ionisation front. This solution can be given as,
\begin{equation}
R_\mathrm{H\&I}(t) = R_\mathrm{st} \left( 1 + \frac{7}{4} \sqrt{\frac{4}{3}} \frac{\sigma_\mathrm{T,i} t}{R_\mathrm{st}} \right) ^{4/7}~,
\end{equation}
This differs from equation\,\ref{spitzer} by a factor of $\sqrt{4/3}$, due to the inclusion of the inertia of the shocked gas (see \citealp{bisbas_2015}). This model also only accounts for the thermal pressure of the ionised gas. \note{We note that the ``early phase'' \citet{spitzer_1978} and \citet{Hosokawa_2006} solutions for thermal expansion have been chosen for their simplicity. However, we have shown that the \HII\ regions studied here could be close to pressure equilibrium with their surrounding environment and, therefore, in a later stage of expansion. We note that there are analytic models that account for the swept-up material in the shell during the later expansion phase(s), which in effect slow the expansion rate for times larger than $\sim$\,0.5\,Myr (e.g. \citealp{raga_2012, williams_2018}). However, we will shortly show that the observed \HII\ region sizes are typically unpredicted by the thermal expansion models. These late time solutions are, therefore, not considered within this section as the further slowing of the expansion rate would not provide a better agreement with the observations.}

A third model for the expansion of an \HII\ region is proposed by \citet{weaver_1977}. This solution accounts for the stellar wind pressure, and takes the following form \citep{tielens_2005}, 
\begin{multline}
R_\mathrm{W}(t) \simeq \left (\frac{2}{\pi} \frac{L_\mathrm{wind}}{\rho_0}t^3 \right )^{1/5}, \\
 \simeq 32 \left ( \frac{L_\mathrm{wind}}{10^{36}\,\mathrm{erg\,s^{-1}}} \right )^{1/5} \left ( \frac{n_\mathrm{cl}}{0.5\, \mathrm{cm^{-3}}} \right )^{-1/5} \left ( \frac{t}{10^6 \mathrm{yr}} \right )^{3/5},
\end{multline}
where $n_\mathrm{cl}$ is the initial molecular hydrogen number density of the host molecular cloud; i.e. $n_\mathrm{cl} = \rho_\mathrm{cl} / \mu_\mathrm{H_2} m_\mathrm{H}$, where $\mu_\mathrm{H_2}=2.8$ is the mean molecular weight per hydrogen molecule, and $m_\mathrm{H}= 1.67 \times 10^{-24}$\,g is the mass of atomic hydrogen. The wind mechanical luminosity, $L_\mathrm{wind}$, is obtained by considering the stellar population within each region. For each high mass star, $L_\mathrm{wind}$ can be expressed as, 
\begin{equation}
L_\mathrm{wind} = \frac{1}{2} \dot{M} v^{2}_\infty~,
\end{equation}
where $\dot{M}$ is the stellar wind mass-loss rate and $v_\infty$ is the terminal wind velocity. Following \citet{mcleod_2019}, we use the mass-loss rates and terminal velocities determined for a range of O-type stellar types from \citet[][Table 1]{muijres_2012}.\footnote{We assume that all the Galactic Centres \HII\ regions contain supergiant stars, and use the last column in \citet[][Table 1]{muijres_2012} for $\dot{M}$, and compute the terminal velocity as 2.6 times the escape velocity (seventh column).} We use the spectral classifications determined from the radio observations outlined in Table\,\ref{radio_obs} to calculate the wind luminosity for the Galactic Centre sources (i.e. \citealp{mehringer_1992, schmiedeke_2016}).  

The final, more complex analytical models we consider were presented by \citet[][section 2, equations 11-13]{krumholzmatzner_2009} and \citet[][section 2, equations 1-5, 10, and 13-16]{kim_2016}. These authors build upon the aforementioned solutions and account for both the thermal and radiation pressure contributions to the expansion. Moreover, \citet[][section 3]{krumholzmatzner_2009} include trapping effects, where the feedback from the young stars is contained within the \HII\ region bubble. The over-pressure caused by the trapped energy results in a significantly increased expansion speed of the \HII\ region.

\begin{figure}
\centering
\includegraphics[trim = 0mm 0mm 0mm 0mm, clip,angle=0,width=1.0\columnwidth]{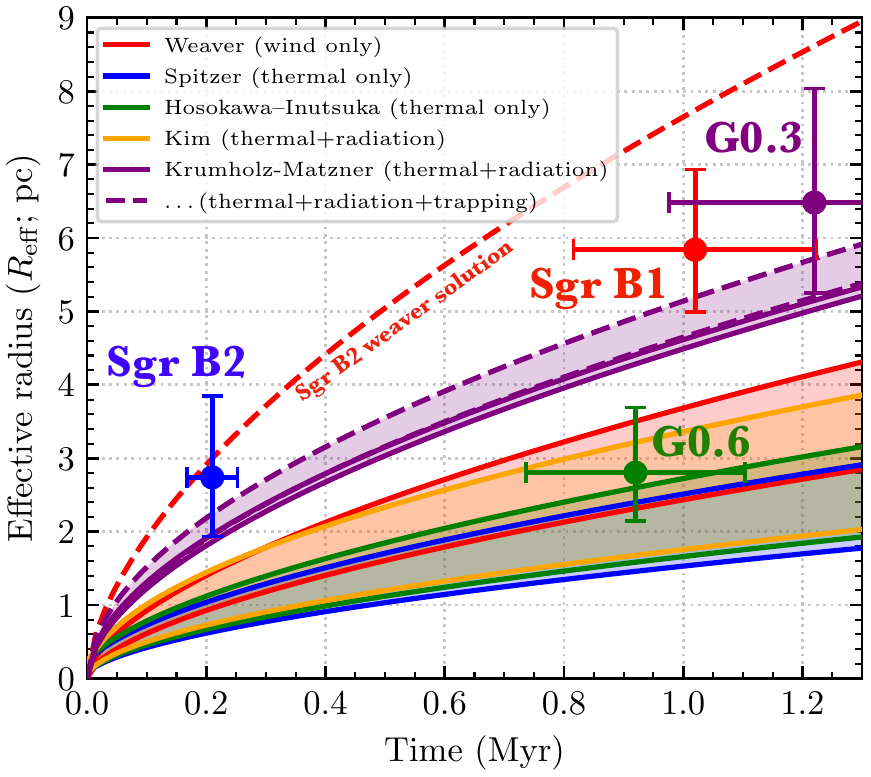}
\caption{A comparison between the sizes of the observed Galactic Centre \ion{H}{II} regions, and the predictions from theoretical models as a function of time (section\,\ref{sec:comp2analmodels}). The model predictions have been taken from \citet{weaver_1977}, \citet{spitzer_1978}, \citet{Hosokawa_2006}, \citet{krumholzmatzner_2009}, and \citet{kim_2016}. These are shown as coloured shaded regions and lines (see legend). The shaded region represents the range of predicted effective radius at a given time within the observational limits for the observed initial cloud properties ($\rho_\mathrm{cl}$, $r_\mathrm{cl}$, $n_\mathrm{cl}$) and \HII\ region properties ($\mathcal{N}_\mathrm{LyC}$, $L_\mathrm{bol}$, $L_\mathrm{wind}$, $f_\mathrm{trap}$). The dashed line shows the \citet{weaver_1977} wind solution using $L_\mathrm{wind}$ determined for the most massive \HII\ region; Sgr B2 (see text). We assume a $\pm20\%$ uncertainty on the age of the observed \ion{H}{II} regions, and calculate the radius uncertainty from the semi-major and minor axis of an ellipse placed over the contour used to define each source (as outlined in \citealp{barnes_2017}).}
\label{expansion_models}
\end{figure}

Figure\,\ref{expansion_models} shows the size as a function of age for each of the analytic models estimated using the range of observed properties within the Galactic Centre \HII\ regions and their progenitor clouds. We adopt Lyman continuum ionisation rates of $\mathcal{N}_\mathrm{LyC} = 0.08\times10^{50}$s$^{-1}$ and $5\times10^{50}$s$^{-1}$, which bracket the range of total ionisation rates measured within G0.6 and Sgr B2, respectively. For \citet{krumholzmatzner_2009}, we use bolometric luminosities of $L_\mathrm{bol}\sim10^{6-7}$\,\Lsol, which have been measured towards Sgr B2 and G0.6, respectively (see \citealp{barnes_2017}). For the \citet{weaver_1977} solution, we use wind luminosities of $L_\mathrm{wind} = 10$\,\Lsol, 100\,\Lsol and 4000\,\Lsol, which cover the measured range within G0.6, Sgr B1 and Sgr B2, respectively. For all the models we adopt the same initial cloud density of $\rho_\mathrm{cl} = 650$\,\sol\,pc$^{-3}$, or $n_\mathrm{cl}=9.5\times10^{3}$cm$^{-3}$, which corresponds to the lowest density estimated for the Galactic Centre precursor molecular clouds (i.e. $m_\mathrm{cl}\sim1.5\times10^{4}$\sol\ and $r_\mathrm{cl}\sim1.75$\,pc for Cloud ``b''; \citealp{barnes_2017}). We use $r_\mathrm{cl}$ and $\rho_\mathrm{cl}$ as $r_\mathrm{0}$ and $\rho_\mathrm{0}$ for the \citet{krumholzmatzner_2009} solution, respectively. These values of $r_\mathrm{cl}$, $\rho_\mathrm{cl}$, $n_\mathrm{cl}$, $\mathcal{N}_\mathrm{LyC}$, $L_\mathrm{bol}$, and $L_\mathrm{wind}$ were chosen to give the most representative expansion rates for the Galactic Centre \HII\ regions. Finally, for the \citet{krumholzmatzner_2009} solution,\footnote{In the models of \citet{krumholzmatzner_2009}, we assume a constant initial density profile ($k_\rho=0$), and the case of a blister \HII\ region, or a constant of 1.9$\times10^{-2}$ in their equations\,4 and 5. This constant has been reduced by a factor of 2.2$^2$ to account for the number of free particles per H nucleus \citep{fall_2010}.} we consider two physical scenarios, represented by two different values of the trapping parameter $f_{\rm trap}$. This parameter represents the factor by which the radiation-pressure force is enhanced by the trapping of energy within the expanding shell. Our first scenario is $f_\mathrm{trap}=1$, which corresponds to every emitted photon being absorbed once in the shell and depositing its momentum there before escaping. Our second scenario is $f_{\rm trap} = 3$, which corresponds to a moderate amplification of the radiation force due to additional scattering of the IR photons produced by the initial absorption and re-emission, or due to the added pressure of hot stellar winds; in terms of pressures, $f_{\rm trap} = 3$ corresponds to $P_{\rm IR} = 2 P_{\rm dir}$ or $P_X = 2 P_{\rm dir}$. \citet{krumholzmatzner_2009} suggest that $f_\mathrm{trap}$ values of a few could be typical for a relatively non-porous shell, and this result is consistent within the uncertainties with our measured value of $P_{\rm IR}$.

The corresponding range of predicted effective radii within the observational limits are plotted as a function of time as coloured shaded regions in Figure\,\ref{expansion_models}. Note that the wind solution for Sgr B2 using $L_\mathrm{wind}$ = 4000\,\Lsol\ has been plotted separately as a dashed red line. We compare the models to the observed sizes and ages of the Galactic Centre \HII\ regions (see Table\,\ref{props_ep}). We assume a $\pm20\%$ uncertainty on the age of the observed \ion{H}{II} regions, and calculate the effective radius uncertainty from the semi-major and minor axis of an ellipse placed over the contour used to define each source (as outlined in \citealp{barnes_2017}; also shown on Figure\,\ref{three_col_map}). 

Figure\,\ref{expansion_models} shows that all the expansion solutions do a reasonable job of predicting the sizes of the observed \HII\ regions given that there has been no fine-tuning of the model parameters. The thermal and radiation pressure solutions reproduce the effective radius of G0.6 well. However, they under-predict the radius of Sgr B2, Sgr B1 and G0.3 by a factor of $\sim$3. We see that the wind and trapped radiation solutions match the radius of Sgr B2  \citep{weaver_1977, krumholzmatzner_2009}. Yet, only the \citet{krumholzmatzner_2009} trapped solution also matches the larger radii of Sgr B1 and G0.3.

Bringing these results together suggests that a modest amount of trapping, either in the form of confined stellar winds or confined IR photons, is important at early times in driving the expansion of the \HII\ regions. Wind feedback, however, needs to become weaker at later times ($>0.2$Myr), to explain the large, more evolved \HII\ regions. The \citet{weaver_1977} solution assumes that the wind gas is adiabatic and trapped, so it applies to a bubble that is completely closed and has no cooling. As soon as gas breaks out, or there is significant mixing between hot and cold gas that leads to cooling, the expansion speed will drop below the \citet{weaver_1977} solution. There is evidence of such strong energy loss from winds for porous \HII\ regions outside of the Galactic Centre \citep{harperclark_2009, rosen_2014}. This porosity could be caused by stellar feedback punching holes in the \HII\ shell or expansion into a non-uniform (turbulent) medium, and can occur relatively early within the \HII\ region's lifetime ($<$1\,Myr). It is entirely possible, however, that in the high-density environment of the Galactic Centre, the winds stay contained within the shell longer, which could lead to a more prolonged expansion.

In summary, we propose that Figure\,\ref{expansion_models} shows evidence for an energy-driven phase of expansion, where either photons or hot gas are contained within the \HII\ regions, but this phase must end by the time the \HII\ region is $\sim$\,1\,Myr old. \note{This timescale is shorter than the predicted cloud lifetime in the Galactic Centre environment \citep{jeffreson18}, again implying that stellar feedback is also an important driver of the cloud lifecycle under high-pressure conditions.}

\note{Finally, in this section, we assess if the large scale expansion of these \HII \ regions has indeed stalled, as is suggested by the balance of their internal pressure components with the surrounding Galactic Centre environment (section\,\ref{sec:comp2SMCLMC}). To do so, we follow \citet{bisbas_2015} and calculate the stalling radius, $R_\mathrm{stall}$, as $R_\mathrm{stall, Sp} = R_\mathrm{st} (\sigma_\mathrm{T,i}/\sigma_\mathrm{T,n})^{4/3}$ for the \citet{spitzer_1978} solution, and  $R_\mathrm{stall, H\&I} = R_\mathrm{st} (\sigma_\mathrm{T,i}/\sigma_\mathrm{T,n})^{4/3}(8/3)^{2/3}$ for the \citet{Hosokawa_2006} solution. In these equations, $\sigma_\mathrm{T,i}\,\sim\,8$\,\kms\ and $\sigma_\mathrm{T,n}\,\sim\,\sqrt{T_\mathrm{cl}k_\mathrm{B}/m_\mathrm{H} \mu_\mathrm{H_2}}\,\sim\,0.4$\,\kms\ are the sound speed in the ionised and neutral gas, respectively (where $\mu_\mathrm{H_2}=2.37$). Using the same parameter ranges assumed above, we calculate $R_\mathrm{stall, Sp}= 4 - 12$\,pc and $R_\mathrm{stall, H\&I} = 8 - 23$\,pc. These cover the measured size range of $\sim\, 2-7$\,pc for the Galactic Centre \HII\ regions (see Figure\,\ref{expansion_models}). Therefore, this analysis further supports that the Galactic Centre \HII\ regions, particularly the largest of these (Sgr B1 and G0.3), may have reached pressure equilibrium with their surrounding environment, and could, therefore, have now stopped expanding.} 

\subsection{Comparison to feedback simulations}\label{simulations}

\begin{figure*}
\centering
\includegraphics[trim = 0mm 0mm 0mm 0mm, clip,angle=0,width=1.0\textwidth]{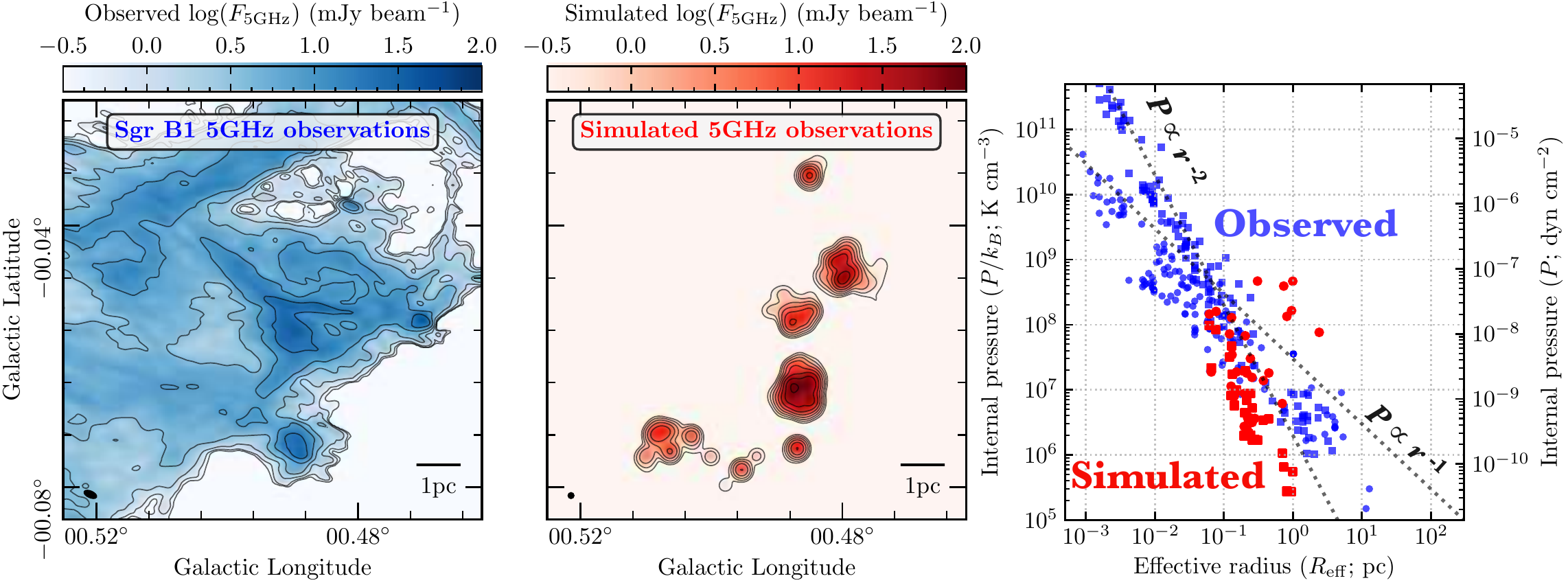}
\caption{A comparison between a Galactic Centre \HII\ region and mock observations taken from the feedback simulations (see section\,\ref{simulations}). The left panel shows a 5\,GHz continuum map taken with the VLA towards the Sgr B1 region (Butterfield et al. in prep), while the central panel is a snapshot from the simulations at a time of $\sim$\,0.8\,Myr, approximately corresponding to the assumed age of Sgr B1 (see Table\,\ref{source_properties}). The simulations have been converted from emission measure units to units of flux densities following \citet{mezger_1967}. To allow a direct comparison to the observations, the colour scale limits are the same for both maps. Contours show isosurfaces of $\log(F_{\rm 5 GHz})$ at intervals of 0.35 dex; the lowest contour corresponds to $10^{-0.5}$ mJy\,beam$^{-1}$, and the highest to $10^2$\,mJy\,beam$^{-1}$. The right panel shows a comparison between the observed and simulated internal pressure components as a function of the \HII\ region effective radius. Here, only the warm ionised thermal pressure ($P_\mathrm{HII}$; circles) and the direct radiation pressure ($P_\mathrm{dir}$; squares) are shown. Note that for the simulated pressures, the largest values were obtained by running our source identification and analysis on spatially smoothed versions of the simulated map shown in the central panel. In doing so, these simulated values offer a direct comparison to the method used to analyse the observed \HII\ regions at increasing (or lower) resolutions within each source. The overlaid diagonal lines shows $P\,\propto\,r^{-1}$ and $P\,\propto\,r^{-2}$ for reference (note that these are not fits to the data).}
\label{simulation_comp}
\end{figure*}

In this section, we compare the observed Galactic Centre \HII\ regions to a set of feedback simulations to better understand the evolution of internal pressure components. To do so, we make use of the smoothed--particle hydro-dynamics simulations described in \citet{dale_2012}, specifically their Run F simulation. Run F models a sub--virial turbulent $10^{5}$\,M$_{\odot}$ cloud with an initial radius of 10\,pc. Such a massive and dense cloud in principle offers a realistic comparison with typical Galactic Centre clouds.

The simulations presented in \citet{dale_2012} allow star formation to initiate, and then model the effects of photoionisation feedback to examine the dynamical effects of the expansion of the \HII\ regions. Star formation occurs at many different locations within the cloud, resulting  in several distinct sources of ionising radiation.

We choose to focus on the snapshot of the Run F simulations at 0.78\,Myr after the formation of the first O--stars and the initiation of ionisation feedback, as this approximately corresponds to the average age of the Galactic Centre \HII\ regions ($\sim$\,0.8\,Myr). We extracted the ionised gas from the snapshot and computed a map of the emission measure integrating along the $z$--axis by integrating the electron density. To produce mock observations, we then convert the emission measure values into units of flux density following the conversions of \citet[][ e.g. also see eq. 6 of \citealp{schmiedeke_2016}]{mezger_1967}. We assume a temperature of 10$^{4}$\,K, a source distance of 8.5\,kpc, a beam size of 5\arcsec, and a frequency of 5\,GHz. The latter two of these parameters were chosen such that we can make a direct visual comparison to some example 5\,GHz VLA observations of the Sgr B1 source (Butterfield et al. in prep), which is the closest in estimated age to the simulation snapshot.
Is
Figure\,\ref{simulation_comp} shows the Sgr B1 VLA observations and the simulated observations over the same angular scale, set to the same colour scale and overlaid with equivalent contour levels. Here, we see similarly compact structures within the two maps that have peak fluxes $\sim$\,0.1\,Jy\,beam$^{-1}$ and sizes of $\sim$\,1pc.\footnote{The spatial resolution of both maps is $\sim$\,0.2\,pc, which limits the identification of any ultra-compact \HII\ regions seen in Sgr B2 by e.g. \citet{depree_1998}.} On the large scale, there appears to be significantly more extended emission within the VLA observations. We believe that this may be due the simulated cloud having no external pressure confining it, so that diffuse ionised gas can leak away to large radii and achieve very low densities. It may also be related to the complex environment of the Galactic Center, which makes it difficult to determine what fraction of the diffuse component is physically associated with the compact sources.

To make a direct comparison to the internal pressures within the observed \HII\ regions, we apply the same analysis presented in section\,\ref{sec:Pressure} to the simulated 5\,GHz observations. Specifically, we smooth the simulation map to the same resolution as the data, and then run a dengrodram analysis on it using the same parameters we use on observations.\footnote{The following set of parameters are used for determination of the dendrogram structure from the simulations: {\sc min\_value} = 10\,$\sigma$ $\sim$ 0.04\,mJy\,beam$^{-1}$; {\sc min\_delta} = 10\,$\sigma$; {\sc min\_delta} = 1\,beam $\sim$ 25 pixels.} We then extract the scale-dependent pressures using the same two methods described in sections sections\,\ref{subsec:Warm ionised gas} and \ref{subsec:Direct radiation}.

The right panel of Figure\,\ref{simulation_comp} shows the components of the internal pressure calculated for the Galactic Centre observations and the simulated observations as a function of radius. Again, we find a nice agreement between the simulations and observations. The match is particularly good on the smallest spatial scales ($\sim$\,0.1\,pc), where we have good crossover in spatial scale between the two data sets. Moreover, it is interesting that the simulations recover the anti-correlation of decreasing pressure with increase size-scale seen in the observations, and appear to have the slope between $P\,\propto\,r^{-1}$ and $P\,\propto\,r^{-2}$ (compare to the overplotted dashed lines). 

In the simulations, the ionising sources at this epoch are still embedded in, or at least close to, the dense, cold filaments in which star formation in the cloud is initiated. It is this gas which is being ionised, and recently--ionised gas, therefore, has high densities. However, the ionised gas rapidly expands into lower--density regions of the cloud, and eventually leaves the cloud entirely, forming an ionised flow (i.e. driven by the thermal ionised gas pressure $P_\mathrm{HII}$) with an expansion velocity on the order of $10$\,km\,s$^{-1}$. We suggest that this may be an explanation of the relation between pressure and size-scale seen in the Galactic Centre clouds. They may also be the result of recently--formed massive stars which are still disrupting the small, dense gaseous structures in which most of the star formation in their host clouds is occurring, by driving pc--scale ionised outflows which disperse the ionised gas to large scales and lower densities and pressures.  

\subsection{Energy and momentum budget}\label{energy}

\note{Within} this section, we assess the coupling efficiency between the total energy injected by the stellar population within the Galactic Centre \HII\ regions, and the observed energy and momentum of the \HII\ regions (section\,\ref{sec:energetics}). 

To determine the total energy injection by the stellar population within each \HII\ region, we simply assume that the luminosity emitted over the lifetime of the stellar population within the \HII\ region could be theoretically used to directly drive its expansion. The coupling efficiency in this case would be $\epsilon_E = E_\mathrm{obs} / E_\mathrm{tot}$, where  $E_\mathrm{obs}$ is the observed kinetic energy of the expanding shell (section\,\ref{sec:energetics}), and $E_\mathrm{tot}$ is the total energy,
\begin{equation}
E_\mathrm{tot} = L_\mathrm{bol} t_\mathrm{age}, 
\end{equation}
where $t_\mathrm{age}$ is the age given in Table\,\ref{source_properties}. 

To determine the total $L_\mathrm{bol}$ for Sgr B2, G0.6 and Sgr B1 we use the highest resolution and quality \HII\ region catalogues available for each source within the literature \citep{schmiedeke_2016, mehringer_1992}. We use the stellar parameters of O and early B type stars determined by \citet[][Table 5]{vacca_1996} to convert the catalogued ZAMS type to a bolometric luminosity. We then sum these individual luminosities to get total bolometric luminosities of $10^{6.5}$\Lsol, $10^{5.3}$\Lsol, and $10^{5.7}$\Lsol\ for Sgr B2, G0.6 and Sgr B1, respectively. Given that approximately $L\,\propto\,M^{3.5}$, we do not account for the contribution to the total luminosity from lower mass stars not observed in the radio continuum (i.e. those with $M<8$\,\sol). 

Using the bolometric luminosities in the above equations gives total energies of $10^{45.9}$, $10^{45.3}$, and $10^{45.8}$\,J for Sgr B2, G0.6 and Sgr B1, respectively. Comparing these to the observed energies given in Table\,\ref{props_ep}, we calculate energy coupling efficiencies of $\epsilon_E = E_\mathrm{obs} / E_\mathrm{tot}$ of $7\times10^{-5}$, $15\times10^{-5}$, $24\times10^{-5}$ for Sgr B2, G0.6, and Sgr B1, respectively. These results are summarised in Table\,\ref{tab_tot_eng}. In short, we find that only a very small fraction of the total energy released by the young stellar populations studied here goes into driving the expansion of the \HII\ regions. We expect then that the vast majority of the energy coming out either goes into ionisation or is just starlight that we observe.

We also assess the momentum budget of the embedded stellar populations with respect to their expanding \HII\ regions. To do so, we follow \citet{dekel_2013}, and calculate the momentum efficiency factor, or the ratio of momentum in the expanding shell to the momentum carried by the radiation field. We define this as the dimensionless factor $\psi_W = p_\mathrm{exp} / (L_\mathrm{bol} c t_\mathrm{age})$, which describes the relative importance of either the radiation field ($\psi_W<1$) or the thermal pressure ($\psi_W>1$) in driving the expansion of the \HII\ region. The parameters for this calculation are given in Tables\,\ref{props_ep} and \ref{tab_tot_eng}. We find that $\psi_W \sim $ 4, 30 and 20 for Sgr B2, G0.6 and Sgr B1, respectively (see Table\,\ref{tab_tot_eng}). Therefore, $\psi_W$ is larger than unity for all the Galactic Center \HII\ regions, highlighting that \note{winds cannot provide the energy required for the observed expansion and the regions should be} thermal pressure dominated. This result is in-line with the \HII\ regions having a larger warm ionised gas pressure components compared to direct radiation pressures when measured over similar scales of $\sim$\,1\,pc (see Figure\,\ref{pressure_vs_radius}). Additionally, it is interesting that we find the that the youngest \HII\ region (Sgr B2; 0.2\,Myr) has a significantly smaller momentum efficiency factor than the older \HII\ regions (e.g. Sgr B1; $\sim$\,1\,Myr). This could be evidence to show that \HII\ regions become more thermal pressure dominated, as opposed to radiation pressure dominated, as they expand and evolve. 

\begin{table}
\centering
\caption{Total bolometric luminosity, energy output, energy coupling,$^a$ and the momentum efficiency factor$^b$ for each of the Galactic Centre \HII\ region (section\,\ref{energy}).}
\begin{tabular}{c c c c c c c }
\hline
Source & log($L_\mathrm{bol}$) & log($E_\mathrm{tot}$) & $\epsilon_E$ $^a$ & $\psi_W$ $^b$ \\ 
& (\Lsol) & (J) & ($10^{-5}$) & \\ 
\hline


Sgr B2 & 6.51 & 45.92 & 6.9 & 3.5 \\
G0.6 & 5.26 & 45.31 & 14.7 & 29.5 \\
Sgr B1 & 5.72 & 45.81 & 24.4 & 19.8 \\

\hline
\end{tabular}
\begin{minipage}{0.95\columnwidth}
    \vspace{1mm}
    $^a$ $\epsilon_E = E_\mathrm{exp} / E_\mathrm{tot}$ or \note{fraction} of energy output by embedded stellar population that have driven the expansion of the \HII\ region (see Table\,\ref{props_ep}).\\
    $^b$ $\psi_W = p_\mathrm{exp} / (L_\mathrm{bol} c t_\mathrm{age})$ or the momentum in expanding shell to the momentum carried by the radiation field. \\
    \end{minipage}
\label{tab_tot_eng}
\end{table}

\section{Conclusions}\label{sec_conclusions}

SNe are thought to play a major role in the self-regulation of star formation in galaxies across cosmic time (\citealp{mckee_1977, maclow_2004, klessen_2016}). However, the efficiency with which SNe energy and momentum couples to the local galactic environment strongly depends on the density distribution of the surrounding ISM (see \citealp{girichidis_2016} and references therein). Feedback processes from the pre-SNe stages of high-mass stars play a significant role in determining the environment into which SNe subsequently explode.  Studying these earliest stages of stellar feedback is then crucial to understanding the coupling of SNe to their environment, and hence their contribution to the energy cycle of star formation in galaxies. 
There are observational constraints on the magnitudes and timescales of early stellar feedback in low ISM pressure environments \citep[e.g.][]{kruijssen19a,chevance20,chevance20b}, yet no such constraints exist for more cosmologically typical high ISM pressure environments. 
In this work, we aim to address this by studying the early evolutionary stages (pre-SNe) of stellar feedback within the central $\sim$100\,pc of the Milky Way's Galactic Centre. 

We investigate the dominant pressures within \HII\ regions using the methods outlined by \citet{lopez_2011}, \citet{lopez_2014}, and \citet{mcleod_2019}. These authors calculate the four sources of pressure responsible for the expansion of \HII\ regions as: thermal pressure from the warm ($10^{4}$\,K) ionised gas ($P_\mathrm{HII}$), direct radiation pressure from the luminous stellar population ($P_\mathrm{dir}$), pressure from the photons released by heated dust ($P_\mathrm{IR}$), and thermal pressure from the shock heated ($10^{6}$\,K) X-ray emitting gas ($P_\mathrm{X}$). Here we calculate three of these, $P_\mathrm{HII}$, $P_\mathrm{dir}$, and $P_\mathrm{IR}$ within four large, Galactic Centre \HII\ region complexes (see Figure\,\ref{pressure_vs_radius}); $P_\mathrm{X}$ is unfortunately inaccessible due to the very large extinction and strong foreground toward the Galactic Centre at soft X-ray wavelengths.

We plot the pressure terms as a function of size scale, and find mean radial dependences of $P_\mathrm{HII} \propto R_\mathrm{eff}^{-1}$, and $P_\mathrm{dir} \propto R_\mathrm{eff}^{-1.5}$ (see Figure\,\ref{pressure_vs_radius_idv}). \note{As a result of this radial variation,} $P_\mathrm{dir}$ dominates on the small scales (0.01\,pc), $P_\mathrm{dir} \sim P_\mathrm{HII}$ on the intermediate scales ($0.01 - 0.1$\,pc), and on the large ($>1$\,pc) scales $P_\mathrm{HII} > P_\mathrm{IR} \sim P_\mathrm{X} > P_\mathrm{dir}$ (section\,\ref{sec:Pressure}). Comparing to \HII\ regions within the Large and Small Magellanic Clouds (LMC and SMC, respectively), where the ambient pressure is 2-3 orders of magnitude smaller than the $\sim 10^{7-8}$ K cm$^{-3}$ found in the Galactic Centre, we find that the radius at which \HII\ regions reach pressure balance with their environments is $\sim 2-3$ pc in the Galactic Centre, versus $>100$ pc in the Magellanic Clouds (section\,\ref{sec:comp2SMCLMC}). Given that the maximum sizes of \HII\ regions in the Galactic Centre and LMC/SMC match the radius at which the internal pressure matches the ambient ISM pressure, we suggest that this shows the \HII\ regions sizes are set by the point of pressure equilibrium with the ambient medium (see Figure\,\ref{fig_MCcomp}). 

We also compare our results to the predictions of models for \HII\ region expansion driven by thermal pressure of the ionised gas, radiation pressure including trapping effects, or stellar winds (see Figure\,\ref{expansion_models}). Combining observed sizes with \HII\ region ages estimated from orbital modelling \citep{longmore_2013a, kruijssen_2015}, we find that three of the four \HII\ regions have radii that are best fit by solutions with a moderate amount of boosting by trapped wind or radiation energy. Wind models where the hot gas is purely adiabatic (e.g. \citealp{weaver_1977}) tend to overpredict \HII\ region radii, while those assuming expansion driven solely by direct starlight or warm gas pressure (e.g. \citealp{Hosokawa_2006, kim_2016}) underpredict them. The best fitting results come from the \citet{krumholzmatzner_2009} model with moderate trapping of wind or IR radiation energy ($f_{\rm trap} \sim 3$ in their notation). Consistent with this conclusion, direct measurement of the momentum budget of the expanding shells suggests that they typically carry $\sim 10$ times the momentum of the direct radiation field, again suggesting that modest amounts of trapped energy are boosting the expansion rate.

We also compare to a set of smoothed--particle hydro-dynamics simulations including photoionisation feedback \citep{dale_2012}. Analysis of synthetic radio maps from the simulation show excellent agreement with the small-scale morphology of the observed \HII\ regions as a function of effective radius. In the simulations, the embedded stellar population is ionising small-scale dense gaseous structures from which stars are forming, and then the ionised gas is dispersing to larger scales and lower densities and pressures. This produces a $P\propto r^{-2}$ profile comparable to that seen in the observations, suggesting that this profile may be an imprint of escaping photoionised gas.

In all, we find that the Galactic Centre \HII\ regions are dominated by the direct radiation pressure on only the smallest scales ($<$0.01\,pc), and at all larger scales they appear to dominated by the thermal pressure of the ionised gas ($>$0.01-10\,pc); there is evidence for a modest contribution from trapped IR radiation or hot stellar wind gas early in the expansion, but by significantly less than would be expected for efficient trapping. We see a link between the ages of the \HII\ regions and the relative importance of both the direct and thermal expansion, which suggest that as the \HII\ regions evolve they also become further dominated by the ionised gas thermal pressure. We find that the thermal pressure-driven expansion then reaches a point of pressure equilibrium with the surrounding environment, at which point further expansion is halted. The high ambient pressure within the Galactic Centre then naturally explains the systematically smaller \HII\ regions compared to those observed with in the LMC and SMC.    

\note{In view of the striking similarity between the \HII\ region radii and the radius at which the region pressure drops to the ambient pressure, we hypothesise that star formation proceeds until the gas inflow can be halted, irrespective of the environment. In this case, star formation and feedback self-regulate such that each cloud attains the integrated star formation efficiency required for blowout, which happens when the nascent stellar population can drive the \HII\ region radius to the cloud (or gas disc) scale height. A similar conclusion was recently reached for giant molecular clouds in NGC\,300, which have a mean separation length that closely matches the gas disc scale height, suggesting that their in-plane spacing is set by feedback bubbles breaking out of the disc \citep{kruijssen19a}. Interestingly, the \HII\ region expansion velocities measured across the nearby galaxy population are highly similar to the ones obtained here for the Galactic Centre and have also been attributed to thermal feedback \citep{kruijssen19a,chevance20,chevance20c,mcleod_20}. Together with the present work, these studies provide evidence for feedback-regulated cloud lifecycles, with surprisingly universal characteristics over three orders of magnitude in ambient gas pressure.}

\section*{Acknowledgements}

\note{We would like to thank the referee for their constructive feedback that helped improve the paper.} We would also like to thank Jonathan Henshaw and Jeong-Gyu Kim for their enlightening discussions on the paper, and Natalie Butterfield and collaborators for providing the 5\,GHz VLA continuum observations shown in this work (project ID: 17A-321). ATB and FB would like to acknowledge funding from the European Research Council (ERC) under the European Union's Horizon 2020 research and innovation programme (grant agreement No.726384/Empire). MRK acknowledges support from the Australian Research Council through its Discovery Projects (award DP190101258) and Future Fellowship (award FT180100375) funding schemes, and from the Alexander von Humboldt Foundation through a Humboldt Research Award. JMDK gratefully acknowledges funding from the Deutsche Forschungsgemeinschaft (DFG, German Research Foundation) through an Emmy Noether Research Group (grant number KR4801/1-1) and the DFG Sachbeihilfe (grant number KR4801/2-1), as well as from the European Research Council (ERC) under the European Union's Horizon 2020 research and innovation programme via the ERC Starting Grant MUSTANG (grant agreement number 714907). In this work we use data that were obtained using the Mopra radio telescope, a part of the Australia Telescope National Facility which is funded by the Commonwealth of Australia for operation as a National Facility managed by CSIRO. The University of New South Wales (UNSW) digital filter bank (MOPS) used for the observations with Mopra was provided with support from the Australian Research Council (ARC), UNSW, Sydney and Monash Universities, as well as the CSIRO.

\section*{Data availability}

A full machine-readable version of Table\,\ref{tab:pressurecomps} is available in the online supplementary of this work. The data underlying this article will be shared on reasonable request to the corresponding author.
\bibliographystyle{mnras}
\bibliography{references}

\appendix
\section{Pressure components}\label{appendix}

\note{In this section, we provide the values of the $P_\mathrm{HII}$, $P_\mathrm{dir}$, and $P_\mathrm{IR}$ pressure components determined within the four Galactic Centre \HII\ regions; Sgr B2, G0.6, Sgr B1, and G0.3. These values form the basis of much of the analysis presented within this work, and are plotted as points in Figures\,\ref{pressure_vs_radius}, \ref{pressure_vs_radius_idv}, \ref{fig_MCcomp}, and \ref{simulation_comp}. Table\,\ref{tab:pressurecomps} gives each of the pressure components and the effective radius over which they have been measured either by the literature works (see Table\,\ref{radio_obs}), or from the source identification routine presented in this work (see sections\,\ref{subsec:Direct radiation} and \ref{subsec:Dust reprocessed emission}). Missing values within the table represent where the measurement set is not available to determine the pressure component; e.g. in the case of sources only identified in the high-resolution radio observations taken from the literature. In this table, we provide the dust reprocessed emission pressure using the extinction determined along each line of sight ($P_\mathrm{IR}$), and using a constant visual extinction of $A_\mathrm{V}$\,=\,20\,mag ($P_\mathrm{IR}$(20\,mag); see section\,\ref{subsec:Dust reprocessed emission}). Missing values within these columns can indicate where the extinction is too high to accurately correct within the dust-modelling routine (see section\,\ref{subsec:Dust reprocessed emission}). The full, machine-readable version of Table\,\ref{tab:pressurecomps} can be obtained from the supplementary online material of this work.} 


\begin{table*}
\centering
\caption{\note{The discrete measurements of the pressure calculated from sources identified from various resolution radio data sets that have been taken from the literature (see Table\,\ref{radio_obs}), or from the sources identified within the available infrared observations (see sections\,\ref{subsec:Direct radiation} and \ref{subsec:Dust reprocessed emission}). Given in columns is the source, the source id, the direct radiation pressure ($P_\mathrm{dir}$; section\,\ref{subsec:Direct radiation}), the warm ionised gas pressure ($P_\mathrm{HII}$; section\,\ref{subsec:Warm ionised gas}), the dust reprocessed emission pressure ($P_\mathrm{IR}$; section\,\ref{subsec:Dust reprocessed emission}), and literature reference. Missing values within the table represent where the measurement set is not available to determine the pressure component; e.g. in the case of $P_\mathrm{IR}$ for sources only identified in the high-resolution radio observations taken from the literature. In this table, we quote the dust reprocessed emission pressure using the extinction determined along each line of sight ($P_\mathrm{IR}$), and using a constant visual extinction of $A_\mathrm{V}$\,=\,20\,mag ($P_\mathrm{IR}$(20\,mag); see section\,\ref{subsec:Dust reprocessed emission}). Missing values within these columns can indicate where the extinction is too high to accurately correct within the dust-modelling routine (see section\,\ref{subsec:Dust reprocessed emission}). The full, machine-readable version of this Table can be obtained from the supplementary online material.}}
\begin{tabular}{cccccccc}
\hline
Source & id & $R_\mathrm{eff}$ & log($P_\mathrm{dir}/k_\mathrm{B}$) & log($P_\mathrm{HII}/k_\mathrm{B}$) & log($P_\mathrm{IR}/k_\mathrm{B}$) & log($P_\mathrm{IR}$(20\,mag)/$k_\mathrm{B}$) & reference \\
$\mathrm{}$ &  & $\mathrm{pc}$ & $\mathrm{K\,cm^{-3}}$ & $\mathrm{K\,cm^{-3}}$ & $\mathrm{K\,erg^{-1}}$ & $\mathrm{K\,erg^{-1}}$ &  \\
\hline
SgrB2 & 1 & 1.447 & 6.367 & \dots  & \dots  & \dots & This work \\
SgrB2 & 2 & 3.235 & 6.210 & \dots  & \dots  & 5.126 & This work \\
SgrB2 & 3 & 1.402 & 6.508 & \dots  & \dots  & 5.984 & This work \\
SgrB2 & 4 & 2.322 & 6.573 & \dots  & \dots  & 5.917 & This work \\
SgrB2 & 5 & 1.652 & 6.530 & \dots  & \dots  & 5.376 & This work \\
SgrB2 & 6 & 1.447 & 6.610 & \dots  & \dots  & 5.396 & This work \\
SgrB2 & 2 & 0.038 & 9.054 & 8.158 & \dots & \dots & Schmiedeke et al. (2016) \\
SgrB2 & 3 & 0.025 & 9.559 & 8.609 & \dots & \dots & Schmiedeke et al. (2016) \\
SgrB2 & 4 & 0.007 & 10.067 & 8.582 & \dots & \dots & Schmiedeke et al. (2016) \\
SgrB2 & 5 & 0.010 & 9.877 & 8.533 & \dots & \dots & Schmiedeke et al. (2016) \\
SgrB2 & 6 & 0.010 & 9.856 & 8.501 & \dots & \dots & Schmiedeke et al. (2016) \\
SgrB2 & 7 & 0.017 & 9.583 & 8.439 & \dots & \dots & Schmiedeke et al. (2016) \\
SgrB2 & 8 & 0.009 & 9.958 & 8.615 & \dots & \dots & Schmiedeke et al. (2016) \\
SgrB2 & 9 & 0.007 & 10.105 & 8.667 & \dots & \dots & Schmiedeke et al. (2016) \\
SgrB2 & 10 & 0.029 & 9.267 & 8.305 & \dots & \dots & Schmiedeke et al. (2016) \\
SgrB2 & 11 & 0.021 & 9.530 & 8.493 & \dots & \dots & Schmiedeke et al. (2016) \\
SgrB2 & 12 & 0.038 & 9.226 & 8.362 & \dots & \dots & Schmiedeke et al. (2016) \\
SgrB2 & 13 & 0.004 & 11.226 & 9.822 & \dots & \dots & Schmiedeke et al. (2016) \\
SgrB2 & 14 & 0.002 & 12.004 & 10.260 & \dots & \dots & Schmiedeke et al. (2016) \\
SgrB2 & 15 & 0.003 & 11.418 & 9.920 & \dots & \dots & Schmiedeke et al. (2016) \\
SgrB2 & 16 & 0.001 & 12.277 & 10.350 & \dots & \dots & Schmiedeke et al. (2016) \\
SgrB2 & 17 & 0.004 & 11.140 & 9.736 & \dots & \dots & Schmiedeke et al. (2016) \\
SgrB2 & 18 & 0.004 & 11.141 & 9.920 & \dots & \dots & Schmiedeke et al. (2016) \\
SgrB2 & 19 & 0.002 & 11.609 & 10.248 & \dots & \dots & Schmiedeke et al. (2016) \\
SgrB2 & 20 & 0.002 & 11.705 & 9.963 & \dots & \dots & Schmiedeke et al. (2016) \\
SgrB2 & 21 & 0.002 & 11.472 & 9.884 & \dots & \dots & Schmiedeke et al. (2016) \\
SgrB2 & 22 & 0.002 & 11.329 & 9.910 & \dots & \dots & Schmiedeke et al. (2016) \\
G0.6 & 1 & 1.708 & 6.266 & \dots & \dots & 4.686 & This work \\
G0.6 & 2 & 1.098 & 6.517 & \dots & \dots & 5.997 & This work \\
G0.6 & 3 & 1.402 & 6.566 & \dots & \dots & 5.901 & This work \\
G0.6 & 4 & 0.666 & 6.504 & \dots & 5.901 & 5.437 & This work \\
G0.6 & 5 & 2.959 & 6.683 & \dots & \dots & 5.829 & This work \\
G0.6 & 6 & 1.380 & 6.596 & \dots & \dots & 5.460 & This work \\
G0.6 & 1 & 0.058 & 8.626 & 8.065 & \dots & \dots & Mehringer et al. (1992) \\
G0.6 & 2 & 0.058 & 8.707 & 8.176 & \dots & \dots & Mehringer et al. (1992) \\
G0.6 & 3 & 0.062 & 8.891 & 8.384 & \dots & \dots & Mehringer et al. (1992) \\
G0.6 & 4 & 0.062 & 8.510 & 7.935 & \dots & \dots & Mehringer et al. (1992) \\
SgrB1 & 1 & 1.181 & 6.379 & \dots & \dots & \dots & This work \\
SgrB1 & 2 & 1.690 & 6.913 & \dots & 6.732 & 6.559 & This work \\
SgrB1 & 3 & 0.942 & 6.957 & \dots & 6.695 & 6.404 & This work \\
SgrB1 & 4 & 1.652 & 6.588 & \dots & 6.696 & 5.845 & This work \\
SgrB1 & 5 & 1.039 & 6.931 & \dots & 6.798 & 6.464 & This work \\
SgrB1 & 6 & 0.504 & 6.828 & \dots & 6.334 & 6.051 & This work \\
SgrB1 & 7 & 1.425 & 6.932 & \dots & 6.789 & 6.531 & This work \\
SgrB1 & 1 & 0.058 & 8.311 & 7.618 & \dots & \dots & Mehringer et al. (1992) \\
SgrB1 & 2 & 0.388 & 7.112 & 7.007 & \dots & \dots & Mehringer et al. (1992) \\
SgrB1 & 3 & 0.039 & 8.688 & 7.917 & \dots & \dots & Mehringer et al. (1992) \\
SgrB1 & 4 & 0.078 & 8.348 & 7.838 & \dots & \dots & Mehringer et al. (1992) \\
SgrB1 & 5 & 0.271 & 7.369 & 7.172 & \dots & \dots & Mehringer et al. (1992) \\
SgrB1 & 6 & 0.062 & 8.054 & 7.284 & \dots & \dots & Mehringer et al. (1992) \\
G0.3 & 1 & 1.181 & 6.749 &  & 6.299 & 6.029 & This work \\
G0.3 & 2 & 0.756 & 6.734 &  & 6.648 & 5.978 & This work \\
G0.3 & 3 & 2.336 & 6.727 &  & 6.299 & 6.009 & This work \\
\dots & \dots  & \dots  & \dots  & \dots  & \dots  & \dots  & \dots  \\
\hline

\end{tabular}
\label{tab:pressurecomps}
\end{table*}

\label{lastpage}
\end{document}